\pdfoutput=1
\documentclass[lettersize,journal]{IEEEtran}
\usepackage{amsmath,amsfonts}
\usepackage{algorithmic}
\usepackage{algorithm}
\usepackage{bbm}
\usepackage{array}
\usepackage{marvosym}
\usepackage[caption=false]{subfig}
\usepackage{textcomp}
\usepackage{stfloats}
\usepackage{url}
\usepackage{verbatim}
\usepackage{graphicx}
\usepackage{multirow}
\usepackage{cite}
\usepackage[table]{xcolor} 
\usepackage{tcolorbox}
\usepackage{subcaption}
\tcbuselibrary{theorems}
\newcolumntype{R}{>{\color{red}}c}
\usepackage{booktabs}
\usepackage{colortbl} 
\usepackage[hidelinks]{hyperref}

\newcommand{\responseblack}[1]{\textcolor{black}{#1}}

\usepackage{tikz} 
\definecolor{lime}{HTML}{A6CE39}
\DeclareRobustCommand{\orcidicon}{%
	\begin{tikzpicture}
	\draw[lime, fill=lime] (0,0) 
	circle [radius=0.134] 
	node[white] {{\fontfamily{qag}\selectfont \tiny ID}};    \draw[white, fill=white] (-0.0625,0.095) 
	circle [radius=0.007];    \end{tikzpicture}
	\hspace{-2mm}}
\foreach \x in {A, ..., Z}{%
	\expandafter\xdef\csname orcid\x\endcsname{\noexpand\href{https://orcid.org/\csname orcidauthor\x\endcsname}{\noexpand\orcidicon}}
}

\begin{document}

\title{A Plug-and-play Model-agnostic Embedding Enhancement Approach for Explainable Recommendation}

\author{Yunqi Mi\textsuperscript{\dag}, Boyang Yan\textsuperscript{\dag}, Guoshuai Zhao\textsuperscript{\Letter}, ~\IEEEmembership{Member,~IEEE,} Jialie Shen, ~\IEEEmembership{Senior Member,~IEEE,} and Xueming Qian, ~\IEEEmembership{Member,~IEEE.}
\thanks{\textsuperscript{\dag} These authors contributed equally to this research.}
\thanks{\textsuperscript{\Letter} Corresponding author.}
\thanks{Yunqi Mi, Boyang Yan, and Guoshuai Zhao are with the School of Software Engineering, Xi’an Jiaotong University, Xi’an
710049, China (e-mail: miyunqi@stu.xjtu.edu.cn; yanboyang@stu.xjtu.edu.cn;
guoshuai.zhao@xjtu.edu.cn\responseblack{).}}
\thanks{Jialie Shen is with City, University of London, U.K.(e-mail: jerry.shen@city.ac.uk).}
\thanks{Xueming Qian is with the Ministry of Education Key Laboratory for Intelligent Networks and Network Security, the School of Information and Communication Engineering, and SMILES LAB, Xi’an Jiaotong University, Xi’an
710049, China (e-mail: qianxm@mail.xjtu.edu.cn).}}

\markboth{Journal of \LaTeX\ Class Files,~Vol.~14, No.~8, August~2021}%
{Shell \MakeLowercase{\textit{et al.}}: A Sample Article Using IEEEtran.cls for IEEE Journals}


\maketitle

\begin{abstract}
Existing multimedia recommender systems provide users with suggestions of media by evaluating the similarities, such as games and movies. To enhance the semantics and explainability of \responseblack{embeddings, }it is a consensus to apply additional information (\textit{e.g.}, interactions, contexts, popularity). However, without systematic consideration of representativeness and value, the utility and explainability of embedding drops drastically. Hence, we introduce \textbf{RVRec}, a plug-and-play model-agnostic embedding enhancement approach that can improve both personality and explainability of existing systems. Specifically, we propose a probability-based embedding optimization method that uses a contrastive loss based on negative 2-Wasserstein distance to learn to enhance the representativeness of the embeddings. In addtion, we 
introduce a reweighing method based on multivariate Shapley values strategy to evaluate and explore the value of interactions and embeddings. Extensive experiments on multiple backbone recommenders and real-world datasets show that RVRec can improve the personalization and explainability of existing recommenders, outperforming state-of-the-art baselines.

\end{abstract}

\begin{IEEEkeywords}
Recommender system, explainable recommendation, game theory, distribution modeling.
\end{IEEEkeywords}

\section{Introduction}




\IEEEPARstart{R}{ecently}, multimedia content (\textit{e.g.}, games and movies) on the Internet has become an integral part of human life. To prevent user confusion and improve retrieval efficiency, multimedia recommender systems (RS) demonstrate promising performance \cite{fc1}. These approaches learn the features of users/items, and then calculate their similarity to make recommendations. However, owing to the lack of precise semantics, these methods suffer from poor representativeness and explainability of features \cite{er1}. Under these conditions, it is a consensus to enrich the semantics and explanations of these methods by introducing additional information (\textit{e.g.}, interactions, contexts, popularity) \cite{p1, p2, p3, p5}. \responseblack{Among these, the methods based on historical item sets are considered the most relevant.} \responseblack{The prototypes are defined as average interacted user/item embeddings of each item/user. For example, when an item is interacted with by ten users, its prototype is the mean of those ten user embeddings. And the prototype of a user is the average of the embeddings of all the items it interacts with.} Meanwhile, the content of the interaction itself reveals the reason for suggestions.



However, in real-world scenarios, the utility of the prototype drops drastically. This is mainly because: 1) \textbf{Suboptimal representation}. The serendipity and stochasticity of the observational data of the RS are inevitable \cite{u1, u2, u3, u4}. As a result, the embeddings within the RS are uncertain, undermining the representation of the prototypes. Using Gaussian distribution modelling and mitigating the uncertainty of embedding space to improve the representation of features has demonstrated promising performance \cite{g1, g2, g3}. 
\begin{figure}
	\centering
	\includegraphics[width=0.95\linewidth]{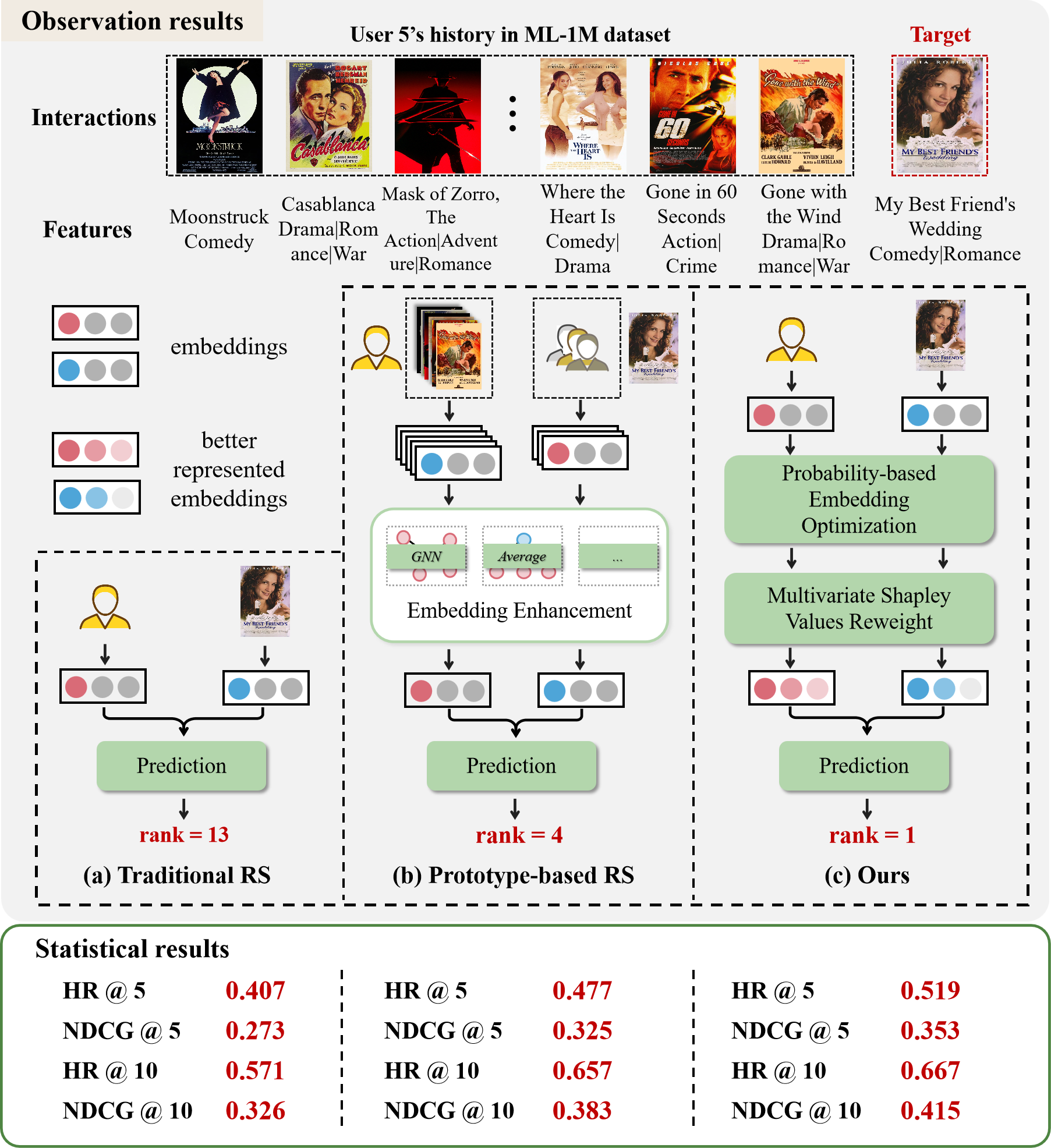}
	\caption{\responseblack{Observation results and statistical results on ML-1M dataset. We use MF~[45] as traditional RS and ProtoMF~[7] as prototype-based RS. We select ProtoMF~[7] as the backbone, so ours means ProtoMF+RVRec. The utility of the prototype drops in real-world scenarios, as shown by the lower rankings in the observation results and poorer performance in the statistical results. }}
	\label{fig:intro}
 \vspace{-0.8cm}
\end{figure} 
 However, due to the smoothing properties and the progressive representation of the tail data points of the Gaussian distribution, they tend to cause fuzzy decision boundaries. Moreover, the existing methods treat embeddings and prototypes as distributions, which are challenging to incorporate effectively into the existing RS architecture. 
2) \textbf{Low-value}. Due to the neglect of the value of different interactions, current RS are suboptimal. Different interaction sets have different contributions to the recommendation result. If some interactions no longer occur, resulting in a significant change to the suggestion, these items/users are regarded as high value \cite{counter1,counter2,counter3}. Current RS focuses on fitting the entire data distribution \cite{counter4,counter5}, which over-learns from low-value interactions and reduces model utility. 
The above issues lead to a decrease in the representative and explainable properties of the prototype, resulting in a suboptimal model.
\responseblack{Figure \ref{fig:intro} shows the observation results and the credible statistical results on the ML-1M dataset. In the experiment, we use MF \cite{mf} as a traditional RS and ProtoMF \cite{p4} as a prototype-based RS. We select ProtoMF \cite{p4} as the backbone, so "Ours" refers to ProtoMF+RVRec. From the observation results, we find that in the traditional RS (MF~\cite{mf}), the target item only ranks \textbf{13} due to its suboptimal representation and low-value features.  Although the prototype-based method (ProtoMF~[7]) improves the ranking (ranks \textbf{4}) by enhancing the feature representation, the performance of the recommendations still needs to be improved. Furthermore, the statistical results of MF and ProtoMF also demonstrate that suboptimal representation and low-value features lead to suboptimal performance of the recommendations. }



\begin{figure*}
	\centering
	\includegraphics[width=1\linewidth]{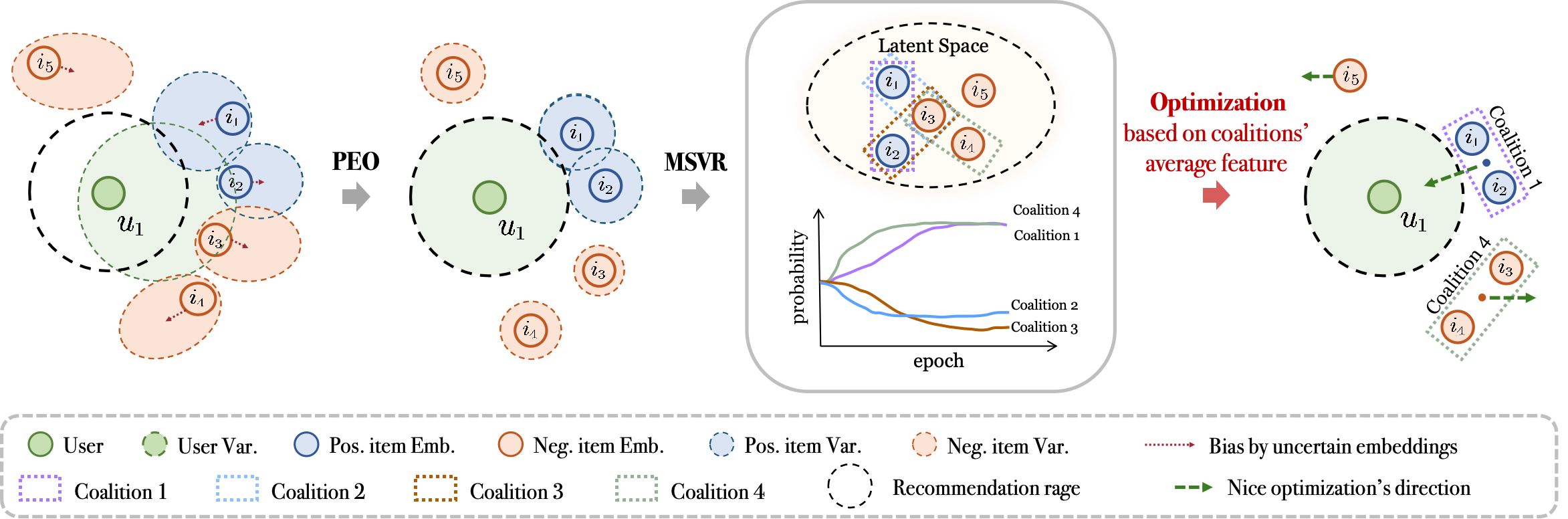}
	\caption{The workflow of RVRec. \textbf{PEO} and \textbf{MSVR} are Probability-based Embedding Optimization module and Multivariate Shapley Values Reweighting module, separately. Among them, PEO aims to optimize the representativeness of embeddings, while MSVR analyzes the interactions between embeddings in a fine-grained way.}
	\label{fig:workflow}
 \vspace{-0.3cm}
\end{figure*} 
To fill this gap, we propose \textbf{RVRec}, a \textbf{r}epresentation and \textbf{v}alue enhanced recommendation method. RVRec introduces a plug-and-play model-agnostic embedding enhancement approach that can pipe into existing RS. Our framework has components: 1) To obtain stable and representative embeddings, we propose a probability-based model for embedding enhancement. It alleviates the uncertainty of embeddings by optimizing the distribution parameters of each user/item. Specifically, we regard the mean value as a more typical embedding of each user/item. We introduce a self-supervised contrastive loss based on the negative 2-Wasserstein distance to improve learning efficiency and prevent dimension collapse. 2) To evaluate and explore the value of interactions, we propose a reweighing method based on multivariate Shapley values. Formally, given a set of user/item interactions, we first initiate the value of each interaction and some coalitions. A coalition's value depends on its members' values, which are optimized during the learning process. However, the computational cost of analyzing all the coalitions' values is unaffordable. Therefore, we suggest a coalition partitioning method based on Bernoulli sampling. To evaluate the value of a coalition, we propose a multiple Shapley value strategy based on data valuation and game theory. To further illustrate, the workflow of RVRec is demonstrated in Figure \ref{fig:workflow}. 
We conduct experiments on three widely used datasets, the results show that RVRec can enhance the representative and value of the prototype, as well as improve recommendation accuracy.

The contributions of our work are as follows: 
\begin{itemize}
     \item 1) We propose a plug-and-play model-agnostic embedding enhancement method RVRec that can improve the representation and value of the features. It can be piped into existing RS directly.
     \item 2) We propose a probability-based embedding optimization method, which can effectively optimize the distribution parameters of each user/item.
     \item 3) We introduce a reweighing method based on multivariate Shapley values strategy to evaluate and explore the value of interactions. To the best of our knowledge, we are the first to explore the value of interactions in RS.
     \item 4) Extensive experiments show that RVRec performs competitively against baselines, with significant improvement both in personalization and explanation.
 \end{itemize}

 \section{Related Work}

\subsection{Explainable recommendation systems} 
Explainable recommendation systems explore how to improve the transparency of algorithms and the reliability of results while making accurate suggestions. The most common stategy is to analyze user behaviour, measuring the similarity between candidate items and interactions for explanation \cite{rw1, rw2, rw3, rw4, rw5}. The insight of these methods is evident, users always interact with similar items. In order to enhance the semantics of embedding and improve explainability, existing methods introduce additional information for RS. The most typical methods rely on interactions, as they respond to the user's real interactions and are usually most relevant to the user \cite{acf, p4, proto1, proto2, proto3}. For instance, ACF \cite{acf} learns a set of anchor embeddings that can represent user and item preferences in a recommender system and generate explanations by exploiting the similarity between embeddings and anchors. ProtoMF \cite{p4} separately define a set of the most typical prototypes of users and items, allowing them to trace their contributions to the predictions and providing direct explanations. However, the lack of comprehensive consideration of embedding representativeness and value sharply drops the utility and explainability of current RS. \responseblack{What's more, current explainable recommendation systems mainly has standalone frameworks, but our method is plug-and-play and model-agnostic, making it highly transferable and easy to integrate with various backbone recommendation models and can provide high plug-and-play model-agnostic explainability. They also neglect the internal relationships and mutual influences between items. The combined
impact of items on user preferences may far exceed the sum of their impacts. We introduce the concept of coalitions to capture how the combination of items impacts user preferences. Hence RVRec can provide more trustworthy and nuanced explanations by considering the collective influence of items.}

\vspace{-0.3cm}
\subsection{Probability Distribution Modeling}
Current machine learning methods commonly represent input data with features (\textit{i.e.},~embeddings) in the latent space, which can lead to bias. This is because there is typically more than one optimal solution to a data representation, \textit{i.e.},~the latent space is uncertain. Modeling the probability distribution of embeddings and the uncertainty of latent space has become a prominent method to solve this issue \cite{pd1, pd2, pd3, pd4}. For example, PFE \cite{pd4} represents each face image as a Gaussian distribution in the latent space to estimate the uncertainty in the latent space for better face classification. DUL \cite{pd1} applys uncertainty learning to face recognition to simultaneously learn features and uncertainty, in addition to analyzing the effects of uncertainty on feature learning. TIGER \cite{pd3} proposes a Gaussian distribution attention method, adaptively aggregating multiple distributions to obtain better word representations. 


Probability distribution modeling stategy also shows promising performance in RS. For instance, DDN \cite{g2} first represents user/item embeddings as Gaussian distributions, effectively mitigating the data sparsity issue. PMLAM \cite{g3} parameterizes each user and item by Gaussian distributions rather than fixed embeddings to capture uncertainties. \responseblack{STOSA and DT4SR \cite{g0, g1} mitigate the cold start issue by introducing uncertainty to model users and their historical item sets to expand users' and items' interaction spaces.} 
\responseblack{DVNE \cite{response1} propose a deep variational embedding framework in Wasserstein space, which captures the uncertainty of embeddings in sparse data scenarios. LVSM \cite{response2} introduce a local variational feature-based similarity model to balance the trade-off between exploration and exploitation by modeling feature uncertainty. DUPLE \cite{response3} propose a dual preference distribution learning framework that models user preferences from item- and attribute-level perspectives. However, their method overlooks the potential influence between items. For example, suppose there are two items from brand A and two items from brand B in the user interactions. Removing both items from brand A may have a significantly greater impact on user preferences compared to removing both items from brand B. We hope to use multivariate shapley value to capture the implicit effect. Furthermore,} existing methods focus on sequence recommendation and cannot apply to other scenarios. In addition, they need to model embeddings as distributions, which cannot fit into the existing RS.

\vspace{-0.3cm}
\subsection{Data valuation}

Data valuation is used in ML to determine a fair compensation for each data owner and identify training examples which have high value \cite{v1,v2}.
In our work, we choose Shapley value, which is commonly used in data valuation. It was originally proposed by Shapley in the field of game theory \cite{game1, game2}. In recent years, there has been an increasing interest in leveraging Shapley values to analyze feature importance and generate counterfactual explanations. \responseblack{WeightedSHAP} \cite{game4} generalize the Shapley value and learn which marginal contributions to focus directly from data. \responseblack{Asymmetric Shapley values} \cite{game5} integrate causal knowledge into Shapley values, providing a more nuanced and context-aware explanation for individual feature contributions in models. \responseblack{Joint Shapley values} \cite{game6} introduce joint Shapley values, which measure the average marginal contribution of a set of features to a model's predictions and naturally extend Shapley's axioms from a single feature to sets of features. \responseblack{Multivariate Shapley interactions in DNNs} \cite{game7} focuses on capturing multivariate Shapley interactions in DNNs, providing a more nuanced and comprehensive interpretation of the model's decision-making process. \responseblack{HBI} \cite{game8} creatively model video-text interactions as game players using multivariate cooperative game theory, proposing Hierarchical Banzhaf Interaction (HBI) to enhance fine-grained semantic interaction and improve cross-modal learning.
\responseblack{However, current RS are suboptimal in the valuation of data, \textit{i.e.}, they neglect the value of different interactions. Furthermore, traditional Shapley value methods cannot be directly applied to recommendation systems due to the need of modeling users and items. To address this, we propose a novel framework that leverages only the embeddings of interactions to compute Shapley values. Based on data valuation and game theory to achieve the evaluation of the value of interactions, we propose multivariate Shapley value for recommender systems, because it provides a principled and rigorous way to evaluate the contribution of coalitions to the recommendation performance. The multivariate Shapley value considers the contributions of items within a coalition, which is crucial for capturing internal relationships among items. }

\section{Methodology}

\subsection{Problem Definition}

Let $ \mathcal{U}=\{\mathbf{u}_{i}\}_{i=1}^{N}$ and $ \mathcal{V}=\{\mathbf{v}_{j}\}_{j=1}^{M}$ represent the user embeddings and item embeddings in a typical RS, respectively. $N$ and $M$ is the number of users and items, and the embeddings are with dimension $d$. \responseblack{In this work, we propose RVRec to improve both recommendation performance and explainability of backbone recommendation models (\textit{e.g.},~MF-based recommenders).} Hence, the user-item co-occurrence matrix $\mathcal{R} \in \mathbb{R} ^{N\times M}$ is binary, where 1 denotes the interaction occurs and 0 vice versa. It should be noted that RVRec is a plug-and-play approach which is not limited to MF-based recommenders. Formally, the goal of the MF-based recommenders is to maximize the predicted probability of real user interactions, which can be defined as follows:
\responseblack{
\begin{equation}\label{eqn:l_rec}
\begin{aligned}
\min \sum_{i\in N}{\mathcal{L}_{\text{REC}}}\left( \theta \right)
&= \min \sum_{i\in N}{\sum_{j\in M}{\ell}(r_{ij},\hat{r}_{ij})} \\
&= \min \sum_{i\in N}{\sum_{j\in M}{\ell}(r_{ij},BM\left( \mathbf{u}_i,\mathbf{v}_j \right) )},
\end{aligned}
\end{equation}
where $\mathcal{L}_{REC}\left(\cdot\right)$ is the error between the prediction $\hat{r}_{ij}$ and the groundtruth $r_{ij}$, $\theta$ is the model parameters, $\ell (\cdot)$ is the specific loss function (\textit{e.g.},~mean squared error or cross-entropy loss), and $BM(\cdot)$ is the backbone recommender, which predicts $\hat{r}_{ij}$ based on the embeddings of the user $\mathbf{u}_i$ and the item $\mathbf{v}_j$. 
}

\begin{figure*}[t]
\centering
\includegraphics[scale=0.228]{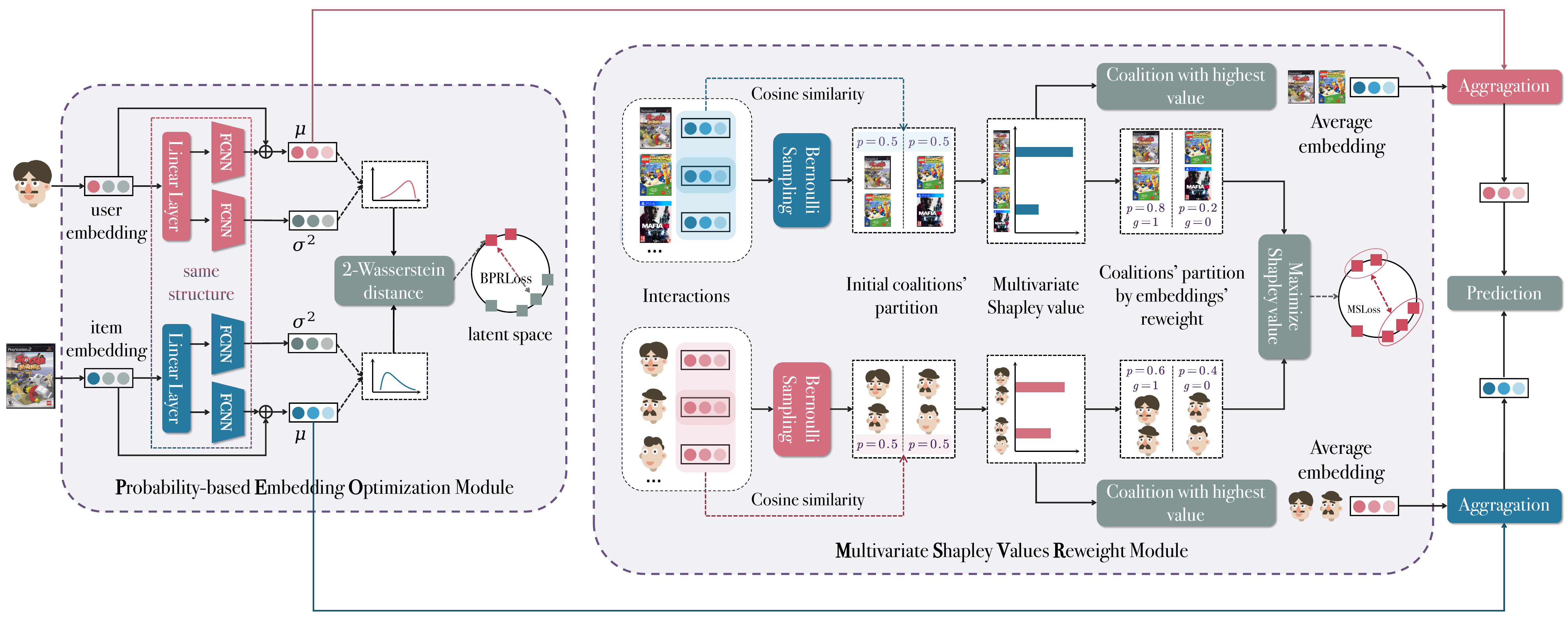}
\caption{The framework of RVRec. \responseblack{RVRec contains two models, probability-based embedding optimization module (PEO) and multivariate Shapley values reweight module (MSVR). PEO is proposed to optimize the distribution of embeddings and improve their representatives. MSVR is proposed to evaluate and explore the value of interactions.}}
\label{fig:1}
 \vspace{-0.5cm}
\end{figure*}

The explanation performance of RVRec comes from the multivariate Shapley values Reweight module. 
\responseblack{We argue that existing methods overlook the potential influence between items. Hence we introduce the concept of \textbf{coalitions} to capture how the combination of items impacts user preferences. The multivariate Shapley values Reweight module's key novelty is that it} measures the contribution of each user/item in the interaction set and the contribution of the coalitions these users/items might form, and reweights the users/items. $\mathcal{R} ^{\mathbf{u}}$ represents users' interactions, $\mathcal{R} ^{\mathbf{v}}$ represents items' interactions. Taking $\mathcal{R} ^{\mathbf{u}}$ as example, each item in $\mathcal{R} ^{\mathbf{u}}$ can be seen as a player. They participate together to contribute to the recommendation result. Different items receive different reward when participating in the game, which is known as the \textit{set function} \cite{game9,game10}, here we use $v\left( \cdot \right)$ to represent it. For the recommendation result, different items have different values, measured by the Shapley value. For the $i$-th item, the Shapley value is:
\begin{equation}\label{equ:2}
\resizebox{0.9\hsize}{!}{$
    \phi \left( i \right) =\boldsymbol{\nu }\left( i \right) =\frac{\left( \left| \mathrm{R}^{\mathbf{u}} \right|-\left| i \right|-1 \right) !\left| i \right|!}{\left| \mathrm{R}^{\mathbf{u}} \right|!} \times \left[ v\left( \mathrm{R}^{\mathbf{u}} \right) -v\left( \mathrm{R}^{\mathbf{u}}/i \right) \right], $}
\end{equation}
when calculating the value of a single item, $\phi \left( i \right)$ is equal to $\boldsymbol{\nu }\left( i \right)$. The specific method for $v\left( \cdot \right)$ is detailed in Equation~\ref{equ:v}. Multiple items can combine a coalition, we use $c_i$ to represent a coalition, design multivariate Shapley value to evaluate the value of $c_i$:
\begin{equation}
\resizebox{0.9\hsize}{!}{$
\begin{aligned}
\phi \left( c_i \right) & = \mathbf{\nu }\left( c_i \right) \times \left| c_i \right|/2+\Sigma _{\mathrm{j}\in c_i}\mathbf{\nu }\left( \mathrm{j} \right) \\
& =\frac{\left( \left| \mathrm{R}^{\mathbf{u}} \right|-\left| c_i \right|-1 \right) !\left| c_i \right|!}{\left| \mathrm{R}^{\mathbf{u}} \right|!}\times \left[ v\left( \mathrm{R}^{\mathbf{u}} \right) -v\left( \mathrm{R}^{\mathbf{u}}/c_i \right) \right] \times \left| c_i \right|/2 \\
& +\Sigma _{\mathrm{j}\in c_i}\frac{\left( \left| \mathrm{R}^{\mathbf{u}} \right|-\left| \mathrm{j} \right|-1 \right) !\left| \mathrm{j} \right|!}{\left| \mathrm{R}^{\mathbf{u}} \right|!}\times \left[ v\left( \mathrm{R}^{\mathbf{u}} \right) -v\left( \mathrm{R}^{\mathbf{u}}/\mathrm{j} \right) \right],
\end{aligned} $}
\end{equation}
where $\left| c_i \right|$ represents the length of $c_i$. 
\responseblack{Next, we use Bernoulli sampling to partition the coalitions. This method identifies which items can form combined coalitions, whose values correspond to the multivariate Shapley value, and which items exist independently, whose values correspond to the Shapley value.}
We use $\Omega$ to represent the partition of coalitions in $\mathcal{R} ^{\mathbf{u}}$, $\mathrm{c}_\mathbf{k}^{*}$ represents all items in $\mathcal{R} ^{\mathbf{u}}$, including coalitions and independent items. The multivariate Shapley values Reweight module's goal is to maximum $\Sigma _{\mathrm{c}_{\mathbf{k}}^{*}\in \Omega}\phi \left( \mathrm{c}_{\mathbf{k}}^{*} \right)$.

\subsection{Framework of RVRec}
Our proposed RVRec framework is shown in Fig.~\ref{fig:1}, which follows the typical MF architecture. Our work concentrates on learning more representative and valuable embeddings for recommenders to improve both personalization and explanation. Thus, we introduce two modules for the MF-based recommenders: probability-based embedding optimization (\textbf{PEO}) and multivariate Shapley values reweight (\textbf{MSVR}). Specifically, the goal of PEO is to optimize the distribution of \responseblack{embeddings} and improve their representatives. We first learn the distribution of embeddings and apply an efficient self-supervised loss for optimization. On the other hand, we use MSVR to evaluate and explore the value of interactions. In detail, given a set of user/item interactions, we design a function $\boldsymbol{\nu }\left( \cdot \right)$ to initiate the value of each interaction and some coalitions. Then, we initiate the coalition partitioning based on the cosine similarity of embeddings and optimize it based on Bernoulli sampling and multiple Shapley values to find users/items that significantly impact the recommendation results. Since RVRec is essentially an improvement to the embedding representation, it can be easily integrated into the recommenders as a plug-and-play module.

\subsection{Probability-based Embedding Optimization}

\subsubsection{Probability Distribution Network}
We firstly detail our proposed PEO strategy, which is mainly composed of a probability distribution network. \responseblack{The inputs of PEO are the user/item embeddings used in recommenders capable of generating user and item embeddings and using them for the final similarity evaluation (\textit{e.g.},~MF-based recommenders, GNN-based recommenders).} For each input, PEO learns a mean vector $\mu$ and variance vector $\sigma ^ 2$, 
\responseblack{which represent the possible distribution center and the variance of the input in the latent space, respectively.} Taking $\mathbf{u_i}$ as an example, we first map the input using a separate linear layer and activation function to obtain a better representation. This process can be formulated as:
\begin{equation}\label{eqn:1} 
    \mathbf{u_i^\ast}=\mathrm{ReLU}\left(\mathbf{W_{\theta_1}}\mathbf{u_i}\right) \in \mathbb{R} ^ d,
    \end{equation}
where $\mathbf{W_{\theta_1}}$ represents the weight of the linear layer. We then calculate the mean and variance of $\mathbf{u_i^\ast}$ using two networks with the same structure\responseblack{, which is widely adopt in recommendation field~\cite{g1, response9}. To decouple their representations, we adopted the idea of ResNet~\cite{response4} and made $\mathbf{u}_{\mathbf{i}}$ serve as an inductive bias to guide the learning of $\mathbf{\mu _{u_i}}$, align base user preferences and dynamic user preferences effectively.} This process can be denoted as:
\begin{equation}\label{eqn:2} 
    \mathbf{\mu _{u_i}}=\mathbf{u_i} + \mathbf{W_{\theta_2}}\mathbf{u_i^\ast},
\end{equation}
\begin{equation}\label{eqn:3} 
    \mathbf{\sigma ^ 2 _{u_i}}=\mathbf{W_{\theta_3}}\mathbf{u_i^\ast},
\end{equation}
where $\mathbf{W_{\theta_2}}$ and $\mathbf{W_{\theta_3}}$ indicate the weights of the two networks, correspondingly. In this paper, we set them as two two-layer fully connected networks. Similarly, we can also learn a distribution $N(\mathbf{\mu _{v_j}}, \mathbf{\sigma ^ 2_{v_j}})$ for the item embedding $\mathbf{v_j}$. The final mean values $\mathbf{\mu _{u_i}}$ and $\mathbf{\mu _{v_j}}$ can be viewed as more representative user/item embeddings than $\mathbf{u_i}$ and $\mathbf{v_j}$ through iteration. For convenience, we still record it as $\mathbf{u_i}$ and $\mathbf{v_j}$ in the following. 

\subsubsection{Probability-based Embedding Optimization}
To improve the distribution accuracy and avoid ambiguous decision boundaries, we apply self-supervised contrastive loss for optimization.
Following STOSA\responseblack{\cite{g1}}, we calculate the similarity between the distributions $N(\mathbf{\mu _{u_i}}, \mathbf{\sigma ^ 2_{u_i}})$ and $N(\mathbf{\mu _{v_j}}, \mathbf{\sigma ^ 2_{v_j}})$ using the negative 2-Wasserstein distance, which can be formulated as:
\begin{equation}\label{eqn:7} 
\resizebox{0.9\hsize}{!}{$
\begin{aligned}
    d\left(\mathbf{u_i}, \mathbf{v_j}\right) = & -\left(d_{W_2}(N(\mathbf{\mu _{u_i}}, \mathbf{\sigma ^2_{u_i}}), N(\mathbf{\mu _{v_j}}, \mathbf{\sigma ^2_{v_j}}))\right)\\
    = & -\left(\left \| \mathbf{\mu _{u_i}} - \mathbf{\mu _{v_j}} \right \| ^2_2 \right.\\
    & \left. + \mathrm{tr} \left(\Sigma _\mathbf{u_i} + \Sigma _\mathbf{v_j} - 2(\Sigma _\mathbf{u_i}^{1/2} \Sigma _\mathbf{v_j} \Sigma _\mathbf{u_i}^{1/2})^{1/2}\right)\right),
\end{aligned}$}
\end{equation}
where $\Sigma$ is the covariance matrix, $d_{W_2}\left(a, b\right)$ is the 2-Wasserstein distance between distribution $a$ and $b$. We choose the 2-Wasserstein distance instead of the KL-divergence because it satisfies the triangular inequality and prevents the vanishing of the gradients, for a detailed proof see \cite{g1,g2}. The negative distance due to the negative correlation between distribution distance and similarity. Actually, $\Sigma$ is the diagonal matrix with the diagonal elements of $\mathbf{\sigma ^2}$. Hence, Eq. \ref{eqn:7} can be denoted as:
\begin{equation}\label{eqn:8} 
    d\left(\mathbf{u_i}, \mathbf{v_j}\right) = -\left(\left \| \mathbf{\mu _{u_i}} - \mathbf{\mu _{v_j}} \right \| ^2_2 + \left \| \mathbf{\sigma _{u_i}} - \mathbf{\sigma _{v_j}} \right \| ^2_2\right).
\end{equation}

Our goal is to maximize the similarity between the interacted user-item pairs while minimizing the similarity between the non-interacted pairs. Hence, we take a contrastive approach by using the Bayesian Personalized Ranking loss function (BPRloss) to measure the similarity gap. The loss of PEO can be formulated as:
\begin{equation}\label{eqn:5} 
\resizebox{0.9\hsize}{!}{$\begin{aligned}
    \mathcal{L}_{PEO} = \displaystyle\sum_{\substack{\mathcal{R}(i, j)=1\\\mathcal{R}(i, k)=0}} \ln(\mathrm{Sigmoid}(d(\mathbf{u_i}, \mathbf{v_j}) - d(\mathbf{u_i}, \mathbf{v_k}))).
    \end{aligned}$}
\end{equation}

In this way, the similarity between distributions is precisely modeled, effectively improving the performance of MF-based recommenders. We show this module in Figure 1, where the backbone network is a prototype-based MF recommender.

\subsection{Multivariate Shapley Values Reweight}

In this section, we describe the proposed multivariate Shapley values reweight in detail. Given a user $\mathbf{u_i}$, $\mathbf{u_i}$'s interactions $\mathcal{R}^{\mathbf{u_i}}$, an item $\mathbf{v_j}$ and $\mathbf{v_j}$'s interactions $\mathcal{R}^{\mathbf{v_j}}$, we get the embeddings of them from Equation~\ref{eqn:2}: $\mathbf{\mu _{u_i}}$, $\mathbf{\mu _{v_k}}\left| \mathbf{k}\in \right. \mathcal{R} ^{\mathbf{u_i}}$, $\mathbf{\mu _{v_j}}$, $\mathbf{\mu _{u_k}}\left| \mathbf{k}\in \right. \mathcal{R} ^{\mathbf{v_j}}$.
We consider each item or user in $\mathcal{R}^{\mathbf{u_i}}$ or $\mathcal{R}^{\mathbf{v_j}}$ as a player, aim to find the coalitions which are more positive. We use $\mathbf{u_i}$ and interacted items $\mathcal{R} ^{\mathbf{u_i}}$ as example. 
When items in $\mathcal{R} ^{\mathbf{u_i}}$ change (e.g., delete), we need a function $ \boldsymbol{\nu }\left( \cdot \right) $ to evaluate the change of the recommendation result to determine the value of the changed items.

For a recommender system, given a user and an item, if the user has interacted with the item, the ground truth is 1, otherwise the ground truth is 0. The recommendation result is evaluated based on the similarity between the inner product of the user and the item's embedding and the ground truth. We consider two adjacent items in $\mathcal{R} ^{\mathbf{u_i}}$ as a coalition, all coalitions in $\mathcal{R} ^{\mathbf{u_i}}$ are defined as:
\begin{equation}
\begin{aligned}
c = \{ & c_1=(\mathbf{v_1},\mathbf{v_2}), c_2=(\mathbf{v_2},\mathbf{v_3}), \ldots, \\
& c_{|\mathcal{R}^{\mathbf{u_i}}|-1}=(\mathbf{v}_{|\mathcal{R}^{\mathbf{u_i}}|-1},\mathbf{v}_{|\mathcal{R}^{\mathbf{u_i}}|}) \},
\end{aligned}
\end{equation}
where $\left| \mathcal{R} ^{\mathbf{u_i}} \right|$ represents the length of $\mathcal{R} ^{\mathbf{u_i}}$. Suppose that the deleted coalition is $c_\mathbf{k}=\left( \mathbf{v_k},\mathbf{v_{k+1}} \right)$, the $ \boldsymbol{\nu }\left( \cdot \right) $ is defined as:
\begin{equation}\label{equ:v}
\begin{aligned}
\boldsymbol{\nu }\left( c_\mathbf{k} \right) = & \frac{\left( |\mathcal{R} ^{\mathbf{u_i}}|-\left( |\mathcal{R} ^{\mathbf{u_i}}|-|c_\mathbf{k}| \right) -1 \right) !\left( |\mathcal{R} ^{\mathbf{u_i}}|-|c_\mathbf{k}| \right) !}{|\mathcal{R} ^{\mathbf{u_i}}|!} \times \\
& \left[ \Sigma _{\mathrm{\mathbf{j}}\in \mathcal{R} ^{\mathbf{u_i}}}\left\| \mathbf{\mu _{u_i}}\cdot \mathbf{\mu _{v_j}}-\mathbbm{1}\left( \mathbf{u_i},\mathbf{v_j} \right) \right\| \right. \\
& \left. - \Sigma _{\mathrm{\mathbf{j}}\in \mathcal{R} ^{\mathbf{u_i}}/c_\mathbf{k}}\left\| \mathbf{\mu _{u_i}}\cdot \mathbf{\mu _{v_j}}-\mathbbm{1}\left( \mathbf{u_i},\mathbf{v_j} \right) \right\| \right],
\end{aligned}
\end{equation}
where $\left\| \mathbf{\mu _{u_i}}\cdot \mathbf{\mu _{v_j}}-\mathbbm{1}\left( \mathbf{u_i},\mathbf{v_j} \right) \right\|$ is the specific method for $v\left( \cdot \right)$ in Equation~\ref{equ:2}, $\left|  c_\mathbf{k} \right|$ is the length of $c_\mathbf{k}$, $\mathcal{R} ^{\mathbf{u_i}}/c_\mathbf{k}$ represents removing $c_\mathbf{k}$ in $\mathcal{R} ^{\mathbf{u_i}}$, $\mathbf{\mu _{u_i}}\cdot \mathbf{\mu _{v_j}}$ is the inner product of $\mathbf{\mu _{u_i}}$ and $\mathbf{\mu _{v_j}}$, $\mathbbm{1}\left( \mathbf{u_i},\mathbf{v_j} \right)$ equals to 1, if $\mathbf{u_i}$ has interacted with $\mathbf{v_j}$, otherwise $\mathbbm{1}\left( \mathbf{u_i},\mathbf{v_j} \right)$ is 0. $\left\| \cdot \right\|$ is the absolute value. Based on $\boldsymbol{\nu }\left( c_\mathbf{k} \right)$, we can calculate the Shapley value of each item in a coalition (\textit{e.g.}, $\mathrm{j}$ in $ c_{\mathbf{k}}$) $\phi \left( \mathrm{j}\in c_{\mathbf{k}} \right)$ and the multivariate Shapley value of coalition (\textit{e.g.}, coalition $ c_{\mathbf{k}}$) $\phi \left( c_{\mathbf{k}} \right)$:

\begin{equation}
\begin{aligned}
\phi \left( \mathrm{j}\in c_{\mathbf{k}} \right) &= \qquad \textcolor{blue}{\boldsymbol{\nu}(c_\mathbf{k})/2}\qquad
 + \textcolor{red}{\boldsymbol{\nu}(\mathrm{j})/2},
\\
\phi \left( c_{\mathbf{k}} \right) &= \textcolor{blue}{\left(\boldsymbol{\nu}(c_\mathbf{k}) \times\left| c_{\mathbf{k}} \right|\right)/2
} + \textcolor{red}{\Sigma_{\mathrm{j} \in c_\mathbf{k}} \boldsymbol{\nu}(\mathrm{j})},
 \end{aligned}
\end{equation}
where $\left| c_{\mathbf{k}} \right|$ represents the length of $ c_{\mathbf{k}}$, $\boldsymbol{\nu }\left( \oslash \right)$ represents the value of items when none of $\mathcal{R} ^{\mathbf{u_i}}$ is deleted. To implement multivariate Shapley, we divide the equation into two parts: the blue part and the red part. The blue part represents the change in the recommendation result when the items in $c_\mathbf{k}$ are both deleted. The red part represents the change in the recommendation result when the items in $c_\mathbf{k}$ are deleted separately. Our goal is to maximize $\phi \left( c_\mathbf{k} \right)$. When the red part is the same and the $\phi \left( c_\mathbf{k} \right)$ is larger, the blue part is larger, which means the coalition is more positive. In addition, if the probability of a coalition's combination is too low (how to calculate the probability of a coalition's combination is explained in Equation~\ref{equ:p}), some items do not belong to any coalition, the Shapley value of these items (here these items are represented by $i$) is calculated by $\phi \left( i \right) =\boldsymbol{\nu }\left( i \right)$.

The computational cost of calculating all the coalitions' shapley value in $\mathcal{R} ^{\mathbf{u_i}}$ is unaffordable, so we use Bernoulli sampling-based method to selectively sample coalitions which have higher probability to combine and be positive, i.e., prototypes which have positive effect on the recommendation result. Finally, the coalition with the greatest value, which is the prototype most likely to represent the user's preferences, is used to enhance the user's embedding.
First of all, for all coalitions $c$ in $\mathcal{R} ^{\mathbf{u_i}}$, we generate the probability of the combination of each coalition $p$. $p$ is calculated by the embeddings' cosine similarity of the items in a coalition, and will be optimized as the embeddings are optimized during training. $p$ is defined as:
\begin{equation}\label{equ:p}
\begin{aligned}
p = \{& p_1 = \cos \left( \mathbf{v_1}, \mathbf{v_2} \right), p_2 = \cos \left( \mathbf{v_2}, \mathbf{v_3} \right), \\
& \ldots, p_{\left| \mathcal{R} ^{\mathbf{u_i}} \right|-1}=\cos \left( v_{|\mathcal{R} ^{\mathbf{u_i}}|-1},v_{|\mathcal{R} ^{\mathbf{u_i}}|} \right)  \},
\end{aligned}
\end{equation}
where $\cos \left( \cdot \right)$ calculates the cosine similarity of two embeddings, $\left| \mathcal{R} ^{\mathbf{u_i}} \right|$ represents the length of $\mathcal{R} ^{\mathbf{u_i}}$. Based on $p$, we use Bernoulli sampling-based method to generate a Bernoulli estimate $g$, defined as:
\begin{equation}
g=\left\{ g_1,g_2,...,g_{\left| \mathcal{R} ^{\mathbf{u_i}} \right|-1} \right\}, 
\end{equation}
where $g_\mathbf{k}\in \left\{ 0,1 \right\}$, $g_\mathbf{k}\sim \mathcal{B} \left( p_{\mathbf{k}} \right)$, $\mathcal{B}$ is Bernoulli estimate function. For example, when $p_\mathbf{k}$ is large, $g_\mathbf{k}$ is more likely to equal to 1, which means $\mathbf{v}_\mathbf{k}$ and $\mathbf{v}_{\left( \mathbf{k+1} \right)}$ combine a coalition $c_\mathbf{k}$, and we calculate the Shapley value  $\phi \left( c_\mathbf{k} \right)$. Otherwise, the coalition will not be combined. We use $\Omega _g$ to denote the partition of coalitions determined by $g$, which is defined as:
\begin{equation}
\Omega _g =\left\{ g_1\times \mathrm{c}_1,g_2\times \mathrm{c}_2,...,g_{\left| \mathcal{R} ^{\mathbf{u_i}} \right|-1}\times c_{|\mathcal{R} ^{\mathbf{u_i}}|-1} \right\}, 
\end{equation}
where $\times$ means multiplication. $\Omega _g$ will be optimized when $p$ is optimized during training.

The complete process of MSVR is to optimize the embeddings of items to obtain $p$, which is the probability of each coalition. Then we use the Bernoulli estimation function $\mathcal{B}$ to get a Bernoulli estimate $g$, obtain the partition of coalitions $\Omega _g$. Finally we calculate the sum of the shapley values of each coalition in $\Omega _g$, defined as $\Sigma _{\mathrm{c}_\mathbf{k}^{*}\in \Omega _g}\phi \left( \mathrm{c}_\mathbf{k}^{*} \right)$. The training goal is to maximize the expectation of $\Sigma _{\mathrm{c}_\mathbf{k}^{*}\in \Omega _g}\phi \left( \mathrm{c}_\mathbf{k}^{*} \right)$, make it closer to 0, i.e., we aim to make the value of the prototype represented by the coalition larger and more closer to the user's preferences. The MSLoss function is defined as:
\begin{equation}
\mathcal{L}_{U-MS}=-\mathbb{E} _{g\sim \mathcal{B} \left( p \right)}\Sigma _{\mathrm{c}_\mathbf{k}^{*}\in \Omega _g}\phi \left( \mathrm{c}_\mathbf{k}^{*} \right),
\end{equation}
where $\mathbb{E} _{g\sim \mathcal{B} \left( p \right)}$ means the expectation of $\Sigma _{\mathrm{c}_\mathbf{k}^{*}\in \Omega _g}\phi \left( \mathrm{c}_\mathbf{k}^{*} \right)$ under the Bernoulli estimate $g$. We also select $\max \left( \phi \left( \mathrm{c}_\mathbf{k}^{*} \right) |\mathrm{c}_\mathbf{k}^{*}\in \Omega _g \right)$, where $\max$ means the maximum value and use $\mathrm{c}_\mathbf{k}^{*}$ as the most important prototype of $\mathbf{u_i}$. We use the average aggregation method $\mathrm{AGG}_{\mathrm{average}}$ to enhance $\mathbf{u_i}$'s embedding by:
\begin{equation}
\mathbf{u}_{\mathbf{i}}^{\mathrm{c}} = \mathrm{AGG}_{\mathrm{average}}\left( \mathbf{\mu _{u_i}},\mathbf{\mu _{v_j}}|\mathbf{j}\in \mathrm{c}_\mathbf{k}^{*} \right).
\end{equation}

Similarly we can get $\mathcal{L}_{I-MS}$, $\mathbf{v}_{\mathbf{j}}^{\mathrm{c}}$ for $\mathbf{v_j}$ and $\mathcal{R} ^{\mathbf{v_j}}$.
The final MSLoss $\mathcal{L}_{MS}=\mathcal{L}_{U-MS}+\mathcal{L}_{I-MS}$. MSVR can be divided into U-MSVR,I-MSVR,UI-MSVR. U-MSVR only generate coalitions for users, I-MSVR only generate coalitions for items. UI-MSVR generate coalitions for users and items.

\subsection{Model Inference and Training}

As mentioned above, RVRec can be directly integrated into existing MF-based recommenders as a plug-and-play component. Therefore, we optimize the prediction results of our method using the recommenders' original prediction loss. The complete loss function for RVRec is formulated as follows:
\begin{equation}\label{eqn:15}
\mathcal{L} = \mathcal{L}_{REC} + \lambda_1 \mathcal{L}_{PEO} + \lambda_2 \mathcal{L}_{MS},
\end{equation}
where $\lambda_1$ and $\lambda_2$ are hyper-parameters. The detailed learning process is as follows. User/item embeddings learned by the feature extractor of the backbone network are used as our input. We use PEO to optimize the embedding representation and obtain more representative features. We then use MSVR to further optimize the embeddings and identify the interrelationships between them. The recommendation reason is a coalition of items that have the greatest impact on user behavior.

\begin{table}[]
\begin{center}
\renewcommand{\arraystretch}{1.3}
\caption{\responseblack{Statistics of the datasets after filtering.}}
\resizebox{\linewidth}{!}{%
\begin{tabular}{lccccc}
\toprule
        Dataset & \# Users   & \# Items & \# Interactions  & Sparsity & Avg. Interactions per User \\ 
\midrule 
   ML-1M     & 6,034   & 3,125     & 574,376      & 96.95\% & 95.23 \\ 
    AMAZON    & 6,950  & 14,494    & 132,209    & 99.87\% & 19.03 \\ 
 LFM2B-1M & 3,555 & 77,985   & 877,365  & 99.68\% & 246.76 \\ 
\bottomrule
\end{tabular}%
}
\label{tab:1} 
\vspace{-0.7cm}
\end{center}
\end{table}

\section{Experiments}
In this section, we conduct extensive experiments on three real-world datasets to demonstrate the validity of RVRec. We focus on the following questions:
\begin{itemize}
    \item \textbf{RQ1}: Does the proposed RVRec improve backbone recommendation models' performance?

    \item \textbf{RQ2}: Does the proposed RVRec improve explanation methods' explanation performance?

    \item \textbf{RQ3}: Whether the individual modules of RVRec (\textit{i.e.}, PEO, MSVR) are valid?

    \item \textbf{RQ4}: What are the influence of hyper-parameters of RVRec?

    \item \responseblack{\textbf{RQ5}: Will the learned representations of RVRec better?}

    \item \responseblack{\textbf{RQ6}: From the perspective of user perception, will RVRec has both the practical utility and explainability?}




\end{itemize}

\begin{table*}[htb]
\centering
\tabcolsep=0.1cm
\scriptsize
\renewcommand{\arraystretch}{1.5}
\caption{The recommendation performance improvement of RVRec on three real-world datasets.}
\begin{tabular}{@{\extracolsep{0pt}}l*{13}{c}}
    \toprule
     \multicolumn{2}{c}{\multirow{2.3}{*}{Recommender}} & \multicolumn{4}{c}{ML-1M} & \multicolumn{4}{c}{AMAZON}& \multicolumn{4}{c}{LFM2B-1M} \\
     \cmidrule(lr){3-6}
     \cmidrule(lr){7-10}
     \cmidrule(lr){11-14}
      & & HR@5 & NDCG@5 & HR@10 & NDCG@10 & HR@5 & NDCG@5 & HR@10 & NDCG@10 & HR@5 & NDCG@5 & HR@10 & NDCG@10 \\
    \midrule
    \multirow{3}{*}{MF} & - & 0.407 & 0.273 & 0.571 & 0.326 & 0.162 & 0.111 & 0.255 & 0.140 & 0.139 & 0.093 & 0.215 & 0.118 \\
    & +RVRec & 0.443 & 0.310 & 0.593 & 0.354 & 0.184 & 0.120 & 0.281 & 0.152 & 0.145 & 0.100 & 0.223 & 0.121 \\
    \rowcolor{red!10}
    & Gains & 8.84\%$\uparrow$ & 13.55\%$\uparrow$ & 3.85\%$\uparrow$ & 8.59\%$\uparrow$ & 13.58\%$\uparrow$ & 8.11\%$\uparrow$ & 10.20\%$\uparrow$ & 8.57\%$\uparrow$ & 4.32\%$\uparrow$ & 7.53\%$\uparrow$ & 3.72\%$\uparrow$ & 2.54\%$\uparrow$ \\
    \midrule
    \multirow{3}{*}{RBMF} & - & 0.346 & 0.230 & 0.505 & 0.282 & 0.112 & 0.075 & 0.166 & 0.093 & 0.288 & 0.208 & 0.384 & 0.279 \\
    & +RVRec & 0.368 & 0.256 & 0.527 & 0.309 & 0.130 & 0.083 & 0.172 & 0.109 & 0.307 & 0.224 & 0.401 & 0.295 \\
    \rowcolor{red!10}
    & Gains & 6.36\%$\uparrow$ & 11.30\%$\uparrow$ & 4.36\%$\uparrow$ & 9.57\%$\uparrow$ & 16.07\%$\uparrow$ & 10.67\%$\uparrow$ & 3.61\%$\uparrow$ & 17.20\%$\uparrow$ & 6.60\%$\uparrow$ & 7.69\%$\uparrow$ & 4.43\%$\uparrow$ & 5.73\%$\uparrow$ \\
    \midrule
    \multirow{3}{*}{ACF} & - & 0.413 & 0.276 & 0.597 & 0.335 & 0.244 & 0.154 & 0.392 & 0.202 & 0.350 & 0.236 & 0.517 & 0.291 \\
    & +RVRec & 0.440 & 0.305 & 0.616 & 0.355 & 0.259 & 0.160 & 0.403 & 0.211 & 0.378 & 0.257 & 0.545 & 0.319 \\
    \rowcolor{red!10}
    & Gains & 6.54\%$\uparrow$ & 10.51\%$\uparrow$ & 3.18\%$\uparrow$ & 5.97\%$\uparrow$ & 6.15\%$\uparrow$ & 3.90\%$\uparrow$ & 2.81\%$\uparrow$ & 4.46\%$\uparrow$ & 8.00\%$\uparrow$ & 8.90\%$\uparrow$ & 5.42\%$\uparrow$ & 9.62\%$\uparrow$ \\
    \midrule
    \multirow{3}{*}{ProtoMF} & - & 0.477 & 0.325 & 0.657 & 0.383 & 0.266 & 0.177 & 0.401 & 0.220 & 0.422 & 0.296 & 0.579 & 0.347 \\
    & +RVRec & 0.519 & 0.353 & 0.667 & 0.415 & 0.293 & 0.198 & 0.434 & 0.245 & 0.438& 0.303 & 0.602 & 0.353 \\
    \rowcolor{red!10}
    & Gains & 8.81\%$\uparrow$ & 8.62\%$\uparrow$ & 1.52\%$\uparrow$ & 8.36\%$\uparrow$ & 10.15\%$\uparrow$ & 11.86\%$\uparrow$ & 8.23\%$\uparrow$ & 11.36\%$\uparrow$ & 3.79\%$\uparrow$ & 2.36\%$\uparrow$ & 3.97\%$\uparrow$ & 1.73\%$\uparrow$ \\
    \midrule
    \multirow{3}{*}{TransGNN} & - & 0.597 & 0.456 & 0.715 & 0.494 & 0.667 & 0.572 & 0.750 & 0.599 & 0.402 & 0.289 & 0.537 & 0.329 \\
    & +RVRec & 0.689 & 0.587 & 0.780 & 0.617 & 0.695 & 0.613 & 0.764 & 0.637 & 0.471 & 0.355 & 0.617 & 0.404 \\
    \rowcolor{red!10}
    & Gains & 15.41\%$\uparrow$ & 28.73\%$\uparrow$ & 9.09\%$\uparrow$ & 24.90\%$\uparrow$ & 4.20\%$\uparrow$ & 7.17\%$\uparrow$ & 1.87\%$\uparrow$ & 6.34\%$\uparrow$ & 17.16\%$\uparrow$ & 22.84\%$\uparrow$ & 14.90\%$\uparrow$ & 22.80\%$\uparrow$ \\
    \midrule
    \rowcolor{red!10}
    \multicolumn{2}{c}{Mean Gains}& 9.19\%$\uparrow$ & 14.54\%$\uparrow$ & 4.40\%$\uparrow$ & 11.48\%$\uparrow$ & 10.03\%$\uparrow$ & 8.34\%$\uparrow$ & 5.34\%$\uparrow$ & 9.59\%$\uparrow$ & 7.97\%$\uparrow$ & 9.86\%$\uparrow$ & 6.49\%$\uparrow$ & 8.48\%$\uparrow$ \\
    \bottomrule
\end{tabular}
\label{tab:rec_result}
\vspace{-0.3cm}
\end{table*}

\vspace{-0.5cm}
\subsection{Experimental Setting}

\subsubsection{Dataset Description}


We conduct experiments on three real-world datasets, including \textbf{MovieLens-1M (ML-1M)} \cite{ml1m}, \textbf{Amazon Video Games (AMAZON)} \cite{AMAZON} and \textbf{LFM2B-1Month (LFM2B-1M)} \cite{lfm2b1mon}. The dataset is described below:

\begin{itemize}
    \item \textbf{ML-1M}: It is a classic movie recommendation dataset that contains 1 million movie ratings on a scale from 1 to 5.
    \item \textbf{AMAZON}: It consists of the ratings on Amazon’s Video Games category on a 1 to 5 scale.
    \item \textbf{LFM2B-1M}: LFM2B consists of users‘ music listening histories, from which extracting one month’s records.
\end{itemize}

Following ProtoMF \cite{p4}, we binarised the dataset since we were focusing on the implicit feedback recommendation task. \responseblack{We fully randomized the training data to ensure that no sequential information is retained.} For ML-1M and AMAZON, we consider ratings above 3.5 as positives and otherwise as negatives. For LFM2B-1M, we further filter the dataset by removing the outlier users that listened to more than the 99$^{\mathrm{th}}$ percentile of all the users, keeping only users with ages between 10 to 95. We use k-core filtering for all datasets, meaning we only keep the users interacting with at least k distinct items and only the items consumed by at least k distinct users. The value of k is 5 for ML-1M and AMAZON and 10 for LFM2B-1M. The statistics of datasets for experiments are describe in table \ref{tab:1}.

\subsubsection{Baselines}
We aim to improve both the personalized recommendation capability and explainability of backbone recommendation models, so we choose the following baselines:

\textbf{Backbone recommendation models.}
\begin{itemize}
    \item \textbf{MF\cite{mf}}. The most typical method to make recommendations by learning the co-occurrence between user-item pairs. 
    \item \textbf{RBMF\cite{rbmf}}. It mitigates the cold-start issue of MF-based recommenders by trade-offs between coverage and diversity of user / item representations. 
    \item \textbf{ACF\cite{acf}}. It learns a set of anchor vectors that simultaneously describe the tastes of both users and items to improve representation semantics. 
    \item \textbf{ProtoMF\cite{p4}}. It uses learnable user / item prototype sets to enhance the representation of embeddings.
    \item \textbf{TransGNN\cite{transgnn}}. It integrates Transformer and GNN layers in an alternating fashion to mutually enhance their capabilities.
\end{itemize}

\textbf{Explanation method.}
\begin{itemize}
    \item \textbf{CCR\cite{ccr}}. It integrates counterfactual and neural logical reasoning to generate counterfactual training examples, improving model accuracy and transparency.
    \item \textbf{Shapley+backbone recommenders}. It combines single-item Shapley value with MF-based recommenders, improving recommendation accuracy and explainability.
\end{itemize}

To further verify the effectiveness of PEO, we select the following probability modeling method for ablation analysis:
\begin{itemize}
    \item \textbf{DDN\cite{g2}}. It models uncertainty using a distribution instead of an embedding vector, which both improves cold start performance and personalization capabilities.
\end{itemize}

\subsubsection{Evaluation Metrics}
We focus on the top-k recommendation task, \textit{i.e.}, whether the top-k candidates in the recommendation list hit the user’s preference. Thus, we evaluate the recommendation performance of our work by two widely used metrics: Normalized Discounted Cumulative Gain at rank $K$ (NDCG@K) and Hit Ratio at rank $K$ (HR@K). This two metrics can be formulated as:
\begin{itemize}
    \item \textbf{HR@K}. The Hit Ratio measures the proportion of true interaction pairs that rank within a specified cutoff number among other negatively sampled interactions with true interaction pairs in the ranked list.
\item \textbf{NDCG@K}. Normalized Discounted Cumulative Gain (NDCG) measures the ranking quality of a list by comparing the relevance of items, with higher relevance scores given more importance at top positions.
\end{itemize}

In addition, to evaluate the persuasiveness of the explanations we generate, we use the sufficiency and necessity metrics presented in \cite{PNPS}. This two metrics can be formulated as:
\begin{itemize}
    \item \textbf{Probability of Necessity (PN)}. This metric calculates the probability that the recommender will \textbf{not} suggest a candidate to the user $u_i$ if a certain interaction $v_j$ is removed (\textit{i.e.}, the item for explanation). This process can be denoted as:
\begin{equation}\label{eqn:9} 
\resizebox{0.8\hsize}{!}{$\begin{aligned}
		\begin{split}
    \rm{PN} = \frac{\sum_{u_i\in U}\sum_{v_j\in R_{i,K}}\rm{PN}_{ij}}{\sum_{u_i\in U}\sum_{v_j\in R_{i,K}}I(E_{ij}\neq \emptyset)}
    , \; \rm{PN_{ij}}=\left \{ 
        \begin{array}{lr}
        1,~\rm{if} \; v_j \notin R_{i,K}^*\\
        0,~\rm{else}\\
        \end{array}
        \right.
\end{split}
\end{aligned}$}
\end{equation}
where $\rm{R_{i,K}}$ is the original top-K recommendation list for user $u_i$. When remove the some certain item(s) $I(\boldsymbol{E}_{ij}\neq \emptyset)$ for $u_i$ and get the new recommendation list $\rm{R_{i,K}^*}$. $I(\boldsymbol{E}_{ij}\neq \emptyset)$ is an identity function: when $\boldsymbol{E}_{ij}\neq \emptyset$, $I(\boldsymbol{E}_{ij}\neq \emptyset)=1$. Otherwise, $I(\boldsymbol{E}_{ij}\neq \emptyset)=0$.

\item \textbf{Probability of Sufficiency (PS)}. Similar to PN, it calculates the probability that the recommender will \textbf{still} suggest a candidate to the user $u_i$ if a certain interaction $v_j$ is removed (\textit{i.e.}, the item for explanation). This process can be denoted as:
\begin{equation}\label{eqn:PS} 
\resizebox{0.8\hsize}{!}{$\begin{aligned}
		\begin{split}
    \rm{PS} = \frac{\sum_{u_i\in U}\sum_{v_j\in R_{i,K}}\rm{PS_{ij}}}{\sum_{u_i\in U}\sum_{v_j\in R_{i,K}}(E_{ij}\neq \emptyset)}
    ,\; \rm{PS_{ij}}=\left \{ 
        \begin{array}{lr}
        1,~\rm{if} \; v_j \in R_{i,K}^{'}\\
        0,~\rm{else}\\
        \end{array}
        \right.
\end{split}
\end{aligned}$}
\end{equation}
where $\rm{R_{i,K}^{'}}$ is the new recommendation list after the intervention is applied, and other notations have similar meanings as above.
\end{itemize}
Meanwhile, we also use their harmonic mean $\rm{F_{NS}} = \frac{2\cdot \rm{PN} \cdot \rm{PS}}{\rm{PN}+\rm{PS}}$ for further experiments.

\subsubsection{Implementation Settings}
We implement our proposed method based on Pytorch. Following ProtoMF \cite{p4}, we employ the leave-one-out strategy \cite{leaveoneout} for all datasets, which means the final two interactions of users are used for validation and test. We use the same policy of ProtoMF \cite{p4}, fixing the number of negative samples (sampled uniformly at random) to 99 for evaluation, while we set it as 20 (sampled based on popularity) for training. We set the training iteration number and the embedding dimension as 100, and the batch size is 512. We use the Adam optimization to train our model, and the hyper-parameters are tuned on the validation dataset. Since the setting of ProtoMF is the same as ours, we follow the baselines reported in it. We carried out each experiment five times for cross-validation and reported the mean results.

\begin{figure*}[t]
	\centering
\subfloat[PN\% on Top-1 and Top-5.]{\includegraphics[width=0.33\textwidth]{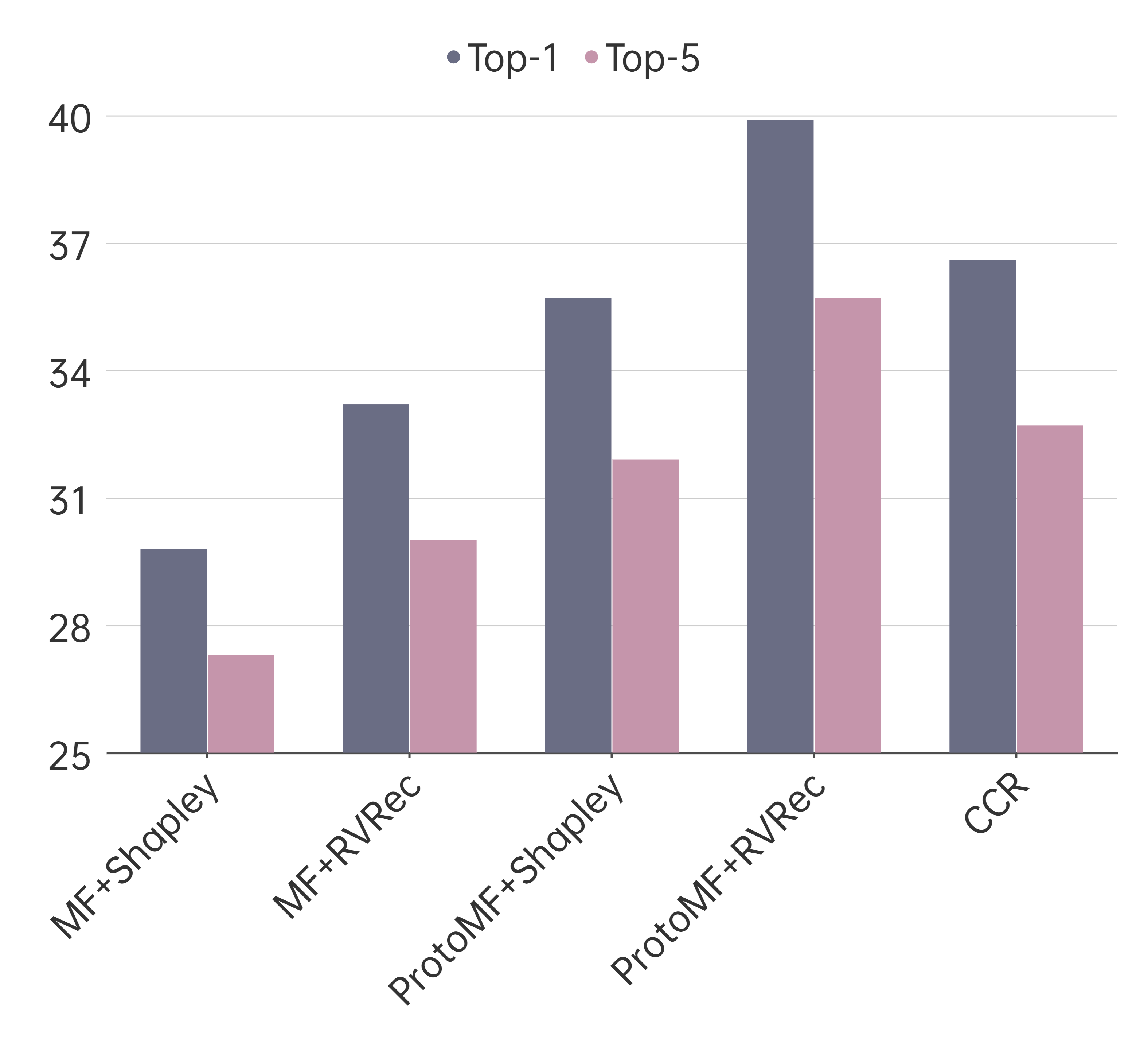}}
 \hfill 	
  \subfloat[PS\% on Top-1 and Top-5.]{\includegraphics[width=0.33\textwidth]{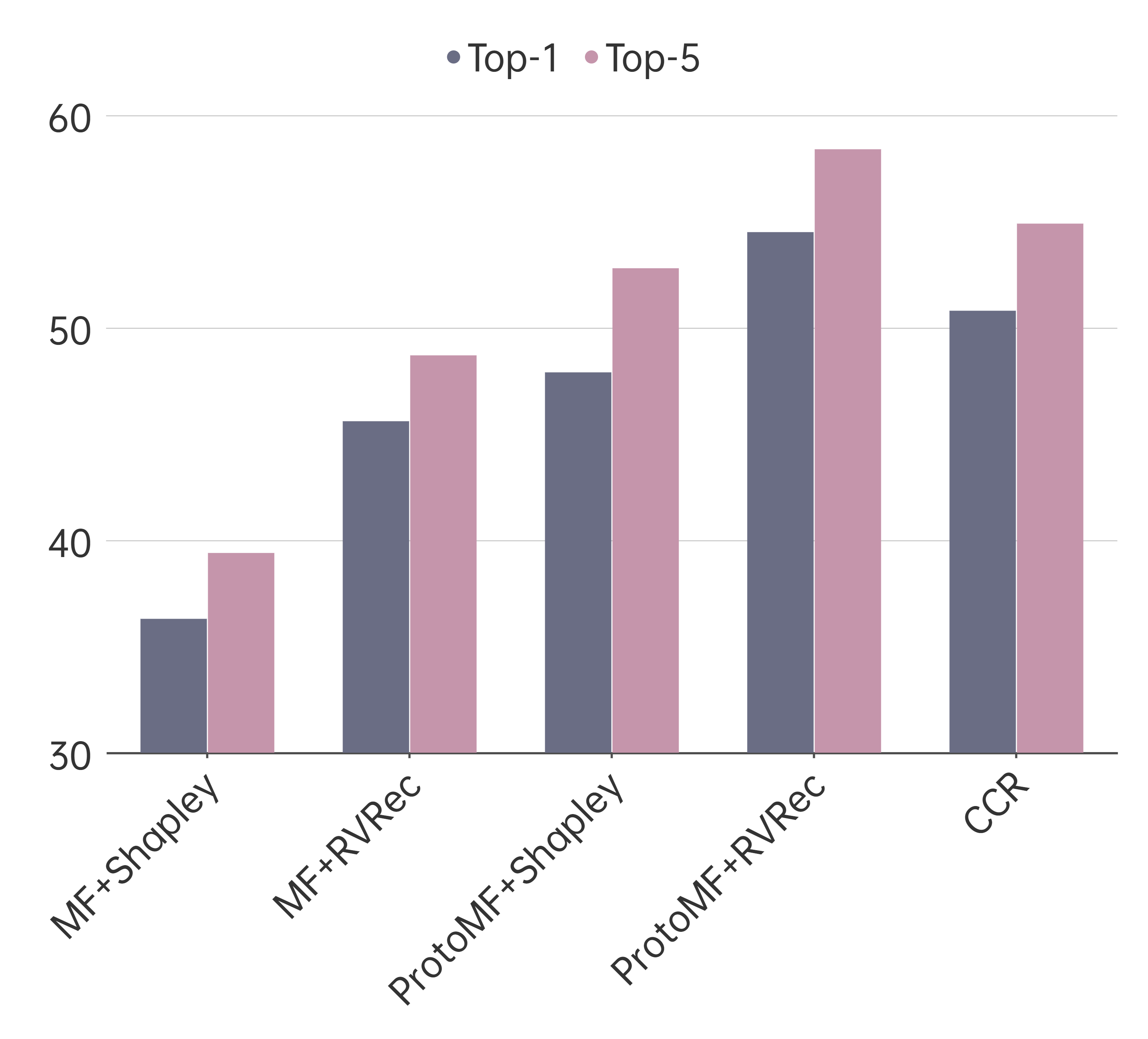}}
  \hfill
  \subfloat[$F_{NS}$\% on Top-1 and Top-5.]{\includegraphics[width=0.33\textwidth]{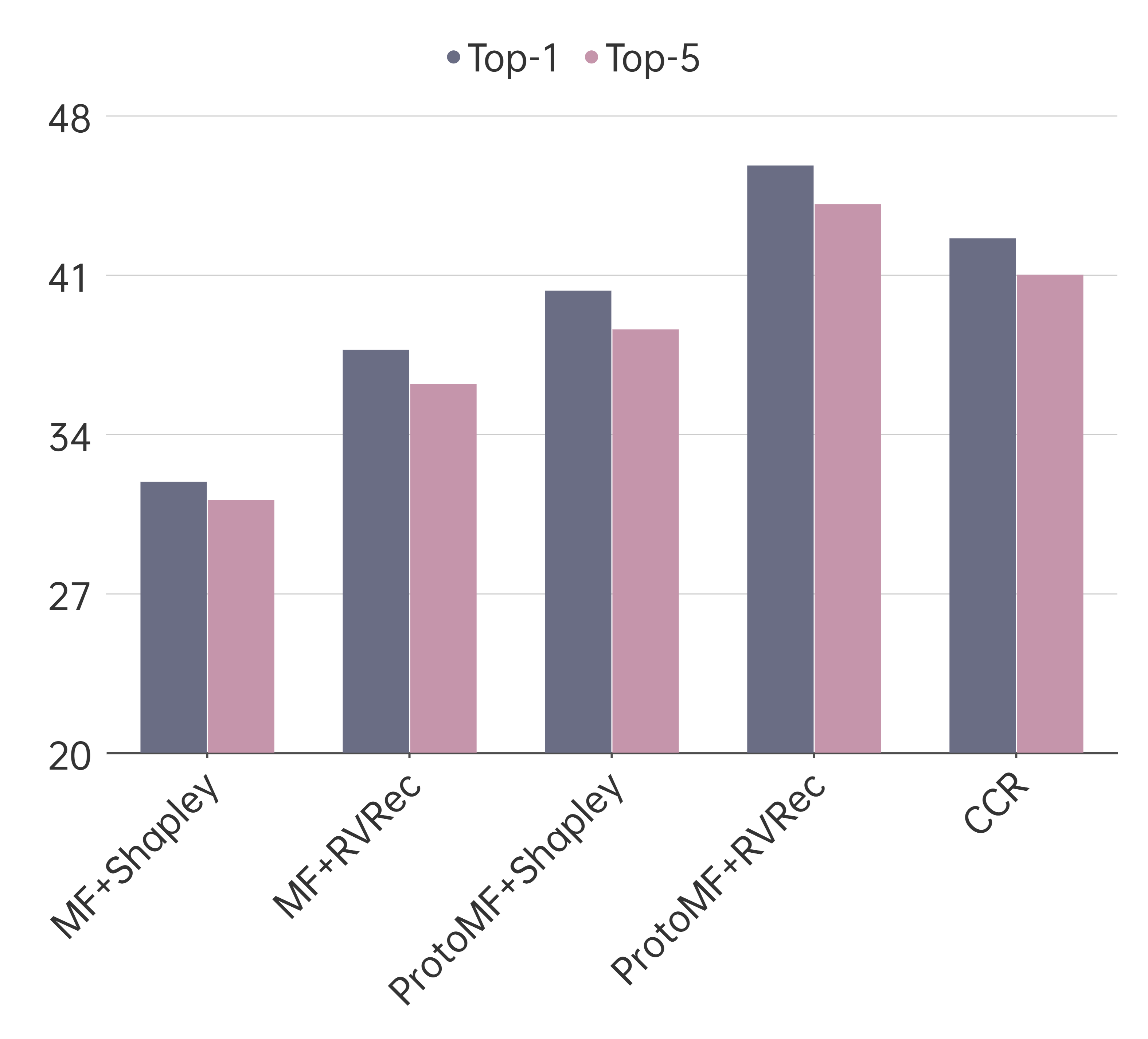}}
	\caption{The explanation performance of RVRec on ML-1M dataset on Top-$K$ ranking task.}
	\label{fig:explanation_compare}
\end{figure*}

\subsection{Recommendation Performance (RQ1)}

This section verifies the personalized recommendation performance of RVRec on three datasets. We plug RVRec into the following recommenders to conduct experiments, including \textbf{MF}, \textbf{RBMF}, \textbf{ACF}, \textbf{ProtoMF} and \textbf{TransGNN}. Specifically, we add RVRec before the embedding crossing and prediction processes of each backbone recommender to learning together. From Table \ref{tab:rec_result}, we have the observation that: 1) \textbf{RVRec can significantly improve the personalised recommendation performance of backbone recommenders.} Experimental results on three real-world datasets show illustrating the superiority of the proposed method. This is mainly because we first effectively model the embedding distribution using a PEO strategy, which significantly improves the semantics and representativeness of the embeddings. Moreover, we use MSVR strategy to analyse the embedding value and its influence on performance, which can accurately identify and analyse the fine-grained similarities between embeddings. 2) \textbf{RVRec is more effective in scenarios with sparse data.} We find that RVRec outperforms on the Amazon dataset than it does on ML-1M dataset, both for the top-5 and top-10 tasks. This shows its outstanding ability to enhance the representation, even in scenarios with sparse data. It is worth noting that the relatively limited improvement on the LFM2B-1M dataset doesn't mean a performance drop. This is because even though this dataset exhibits relatively low sparsity, there are more items (5 times that of the AMAZON dataset) and fewer users (half that of the AMAZON dataset). Considering the popularity bias, there are a large number of items that are not interacted with any user. Even in this scenario, RVRec still shows promising performance, which fully demonstrates the superiority of the proposed method. 
\responseblack{Furthermore, as shown in Table~\ref{tab:1}, the Amazon dataset has a sparsity of \textbf{99.87\%} and an average of only \textbf{19.03} interactions per user, making it highly sparse and exacerbating the user cold-start problem. This poses significant challenges for traditional models (MF, RBMF) to learn effectively. In contrast, TransGNN~\cite{transgnn} leverages Graph Neural Networks (GNNs) to propagate and aggregate information across graph structures, mitigating cold-start problems by inferring user-item relationships through shared neighbours and structural similarities~\cite{response5}. RVRec further enhances these representations, resulting in improved performance.
}

\subsection{Explanation Performance (RQ2)}

This section verifies the explanation performance of RVRec on three datasets. We plug RVRec into the MF and ProtoMF recommender and compare it to the explainable baselines mentioned above separately. Compared to the existing explanation baselines, RVRec can significantly improve the explainability of the prediction results of the backbone recommenders. This is due to the fact that RVRec learns more representative embeddings and analyses the interactions between them. This allows RVRec to more accurately recognize and analyze the essential of user behavior. In contrast to the method using Shapley values, the RVRec-based recommenders has better explainability. It suggests that it is beneficial to introduce Multivatiate Shapley Values for explainable recommender systems.
Overall, the explanation performance on the top-1 ranking task is better than that on the top-5 ranking task. Showing that the accuracy of the recommendation results affects its explainability. This is understandable, as the accuracy of the recommendation is highly dependent on the performance of the RS model's embedding learning and crossing. Meanwhile, the transparency and persuasiveness of the recommendation results also affect the user'trust in the suggestions and the probability of adoption. This indicates a mutually beneficial relationship between explainability and accuracy of recommendations. It is clear that RVRec can improve both, which demonstrates its superiority.

\subsection{Case Study (RQ2)}
\begin{figure}[t]
\centering
\includegraphics[width=1.0\linewidth]{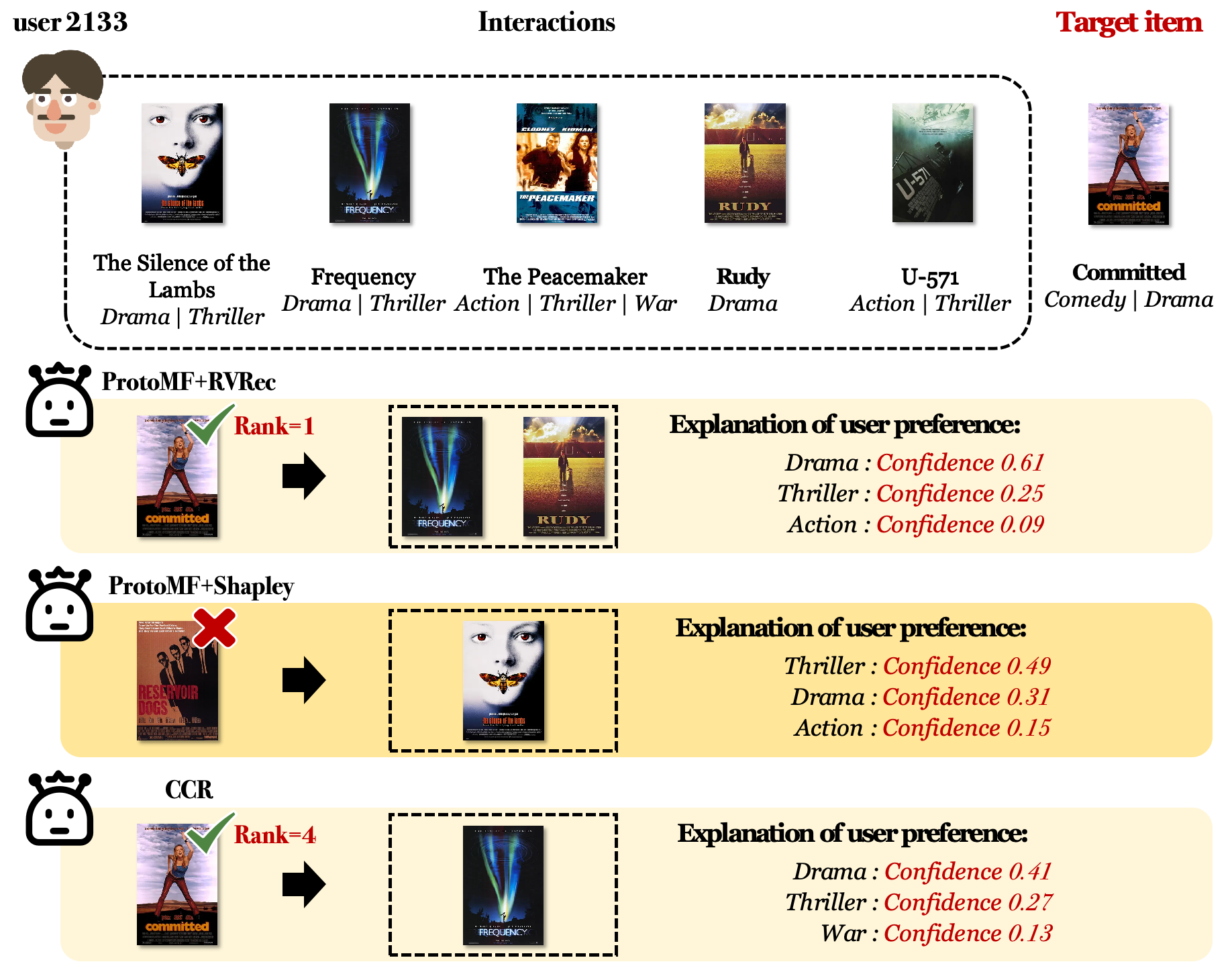}
\caption{Case study on Top-5 ranking task of ML-1M dataset.}
\label{fig:case_study}
\vspace{-0.4cm}
\end{figure}

For further illustration, we use a real user (No. 2133) in ML-1M dataset as an example to comprehensively analyze the performance of different explainable methods. \responseblack{The explanation labels (\textit{e.g.}, Drama, Thriller) are derived from the movie category labels in the dataset. To compute the confidence score, we first obtain the top-5 ranked list of items after model inference. Then, we collect the genre labels for all items in the list and calculate the proportion of each genre label. This process is repeated five times as part of cross-validation, and the mean results are reported.} We select the three best performing methods in the above experiment to evaluate the performance on top-5 ranking task, which are \textbf{ProtoMF+Shapley}, \textbf{CCR} and \textbf{ProtoMF+RVRec}. \responseblack{As shown in \ref{fig:case_study},} for overall performance, ProtoMF+RVRec outperforms the other baselines. It shows that RVRec can effectively alleviate the embedding uncertainty and improve the embedding semantic representation in the RS. The recommendation result of ProtoMF+Shapley are incorrect, indicating that it is not sufficient to simply introduce Shapley values into the backbone recommenders. It further highlights the importance of analyzing the interactions between embeddings. It is understandable to note that the explanation of ProtoMF+RVRec is better than that of CCR. Since ProtoMF+RVRec finds the optimal coalition (including two items) that best supports the user behavior, while CCR only selects one item for explanation. It is clear that RVRec can increase users' trust in the current RS.

\begin{table}[t]
\centering
\renewcommand{\arraystretch}{1.2}
\captionof{table}{\responseblack{The effect of the inductive bias in PEO module on ML-1M dataset with backbones (MF and ProtoMF).}}
\scalebox{0.95}{
\begin{tabular}{ccccc}
\toprule
\multirow{2.5}{*}{Model} & \multicolumn{4}{c}{ML-1M}       \\ \cmidrule{2-5}
& HR@5 & NDCG@5 & HR@10 & NDCG@10 \\ 
\midrule
MF+RVRec(w/o bias)      & 0.436 & 0.305  & 0.589 & 0.350        \\
MF+RVRec                & \textbf{0.443} & \textbf{0.310}  & \textbf{0.593} & \textbf{0.354}        \\ \midrule
ProtoMF+RVRec(w/o bias) & 0.511 & 0.348  & 0.662 & 0.406        \\
ProtoMF+RVRec           & \textbf{0.519} & \textbf{0.353}  & \textbf{0.667} & \textbf{0.415}  \\
\bottomrule
\end{tabular}}
\label{tab:PEObias}
\end{table}

\begin{figure}[t]
	\centering
\subfloat[Effect of PEO.]{\includegraphics[width=0.24\textwidth]{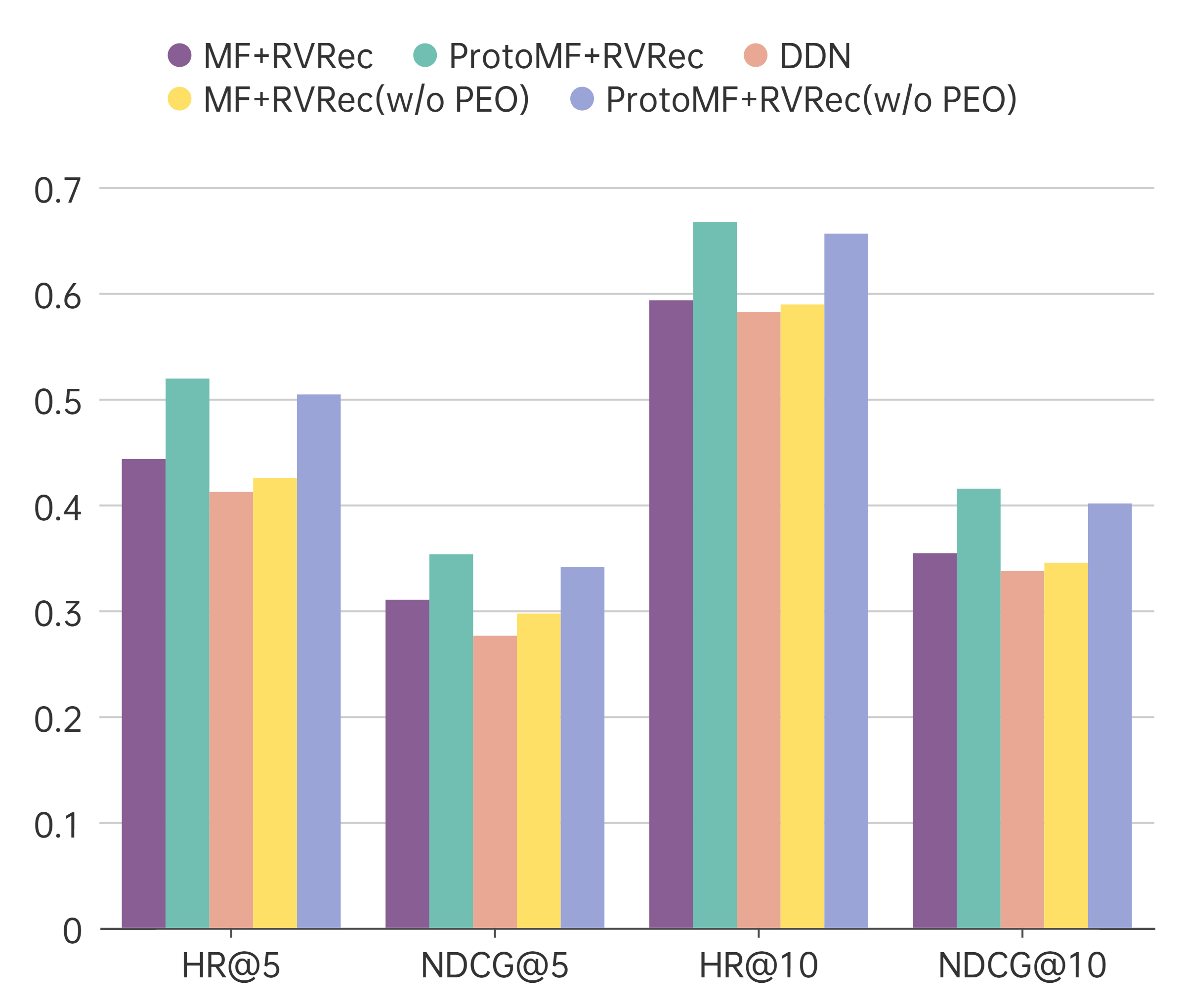}}
 \hfill 	
  \subfloat[Cold-start performance of PEO.]{\includegraphics[width=0.24\textwidth]{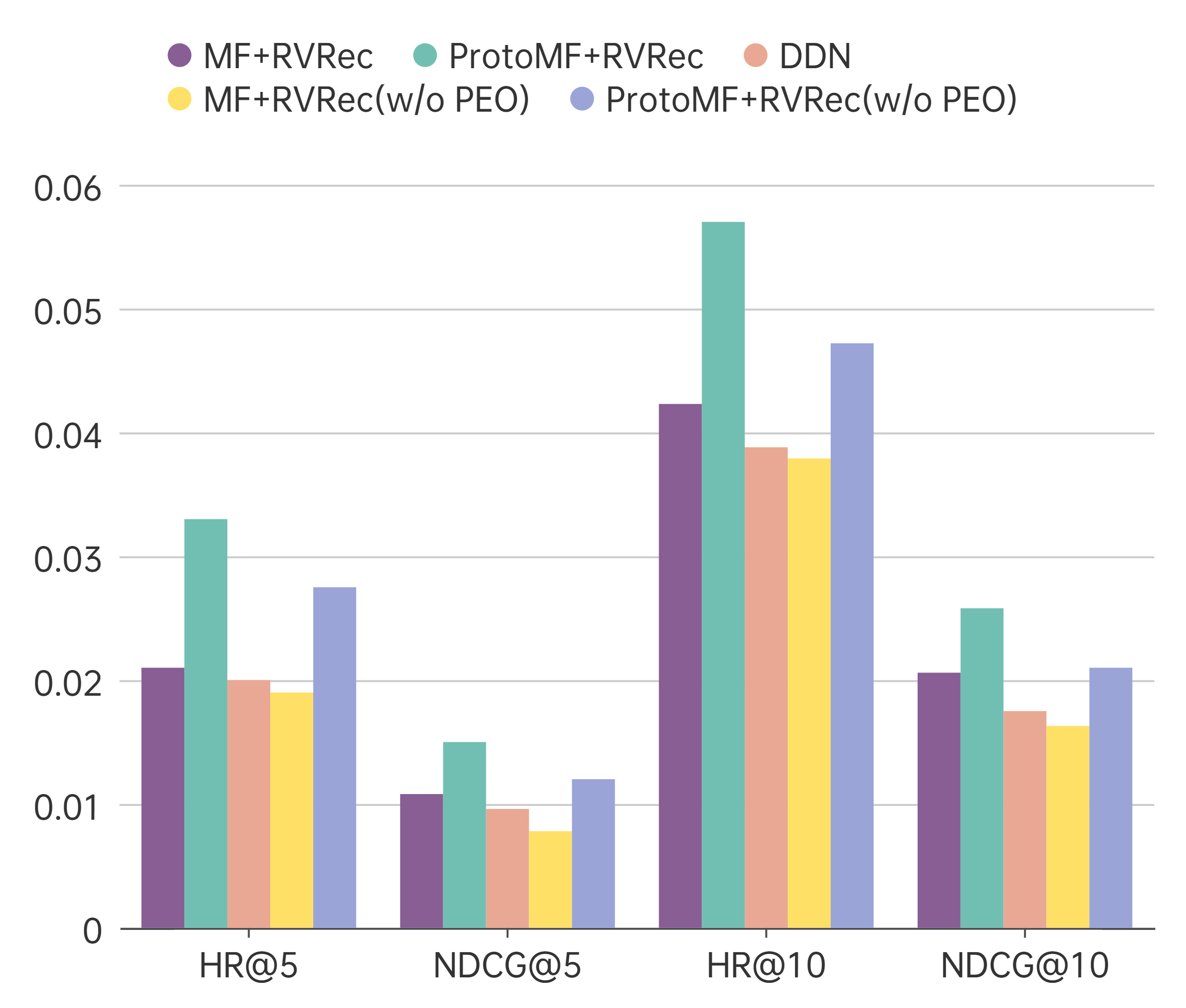}}
	\caption{The effect and cold-start performance of PEO on ML-1M dataset with backbones (MF and ProtoMF).}
	\label{fig:PEO}
 \vspace{-0.2cm}
\end{figure}

\begin{figure}[t]
	\centering
\subfloat[$\lambda_1=1$.]{\includegraphics[width=0.23\textwidth]{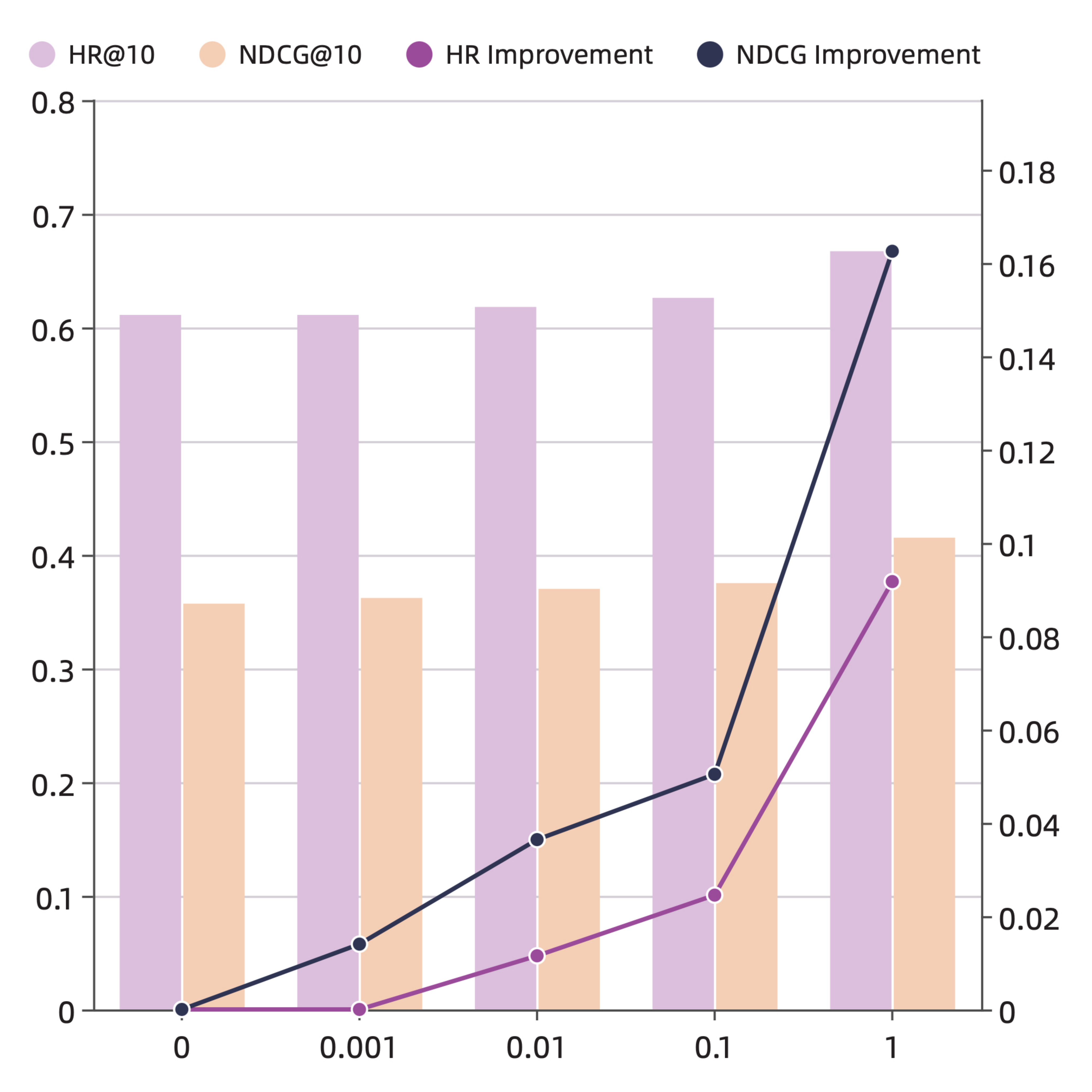}}
 \hfill 	
  \subfloat[$\lambda_2=1$.]{\includegraphics[width=0.23\textwidth]{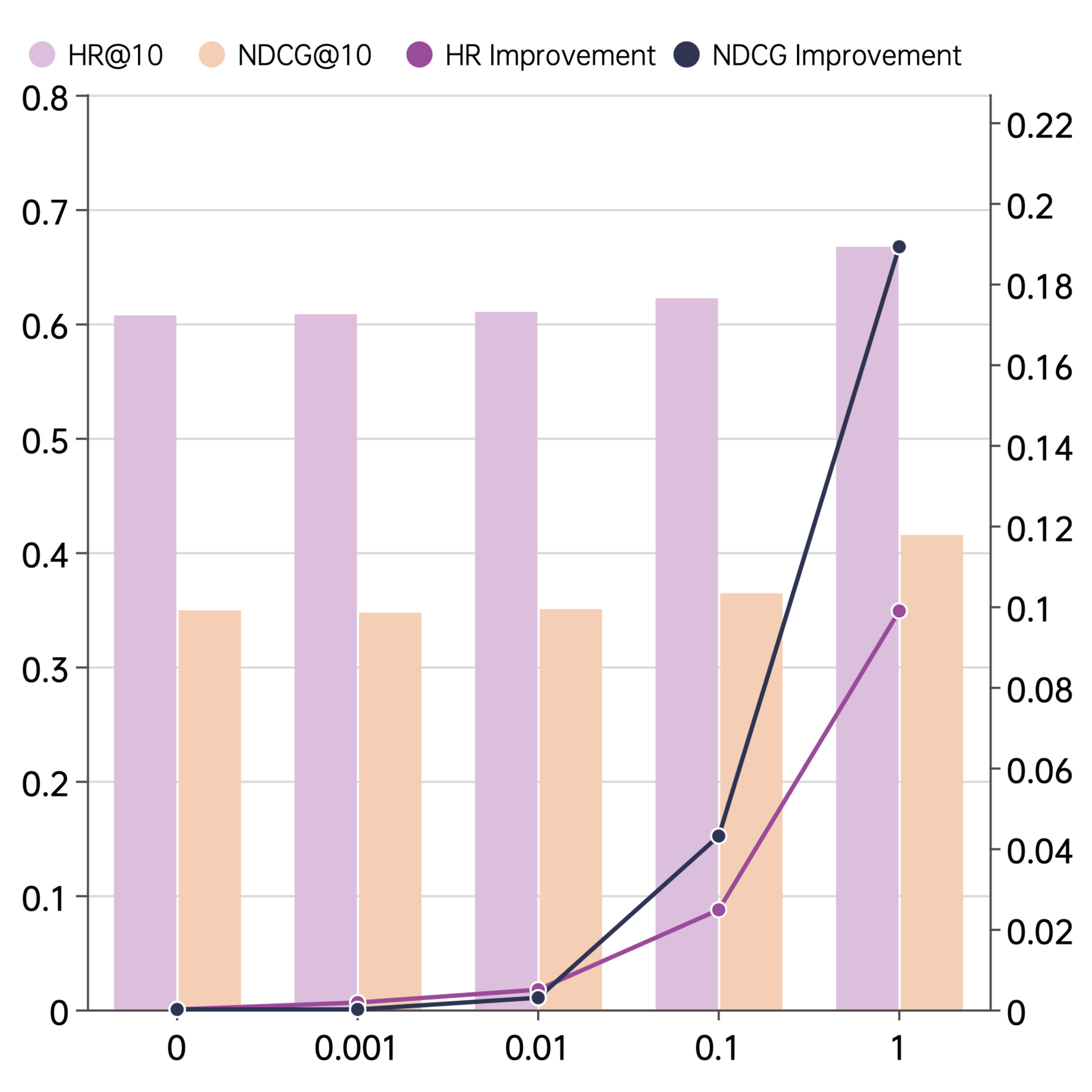}}
	\caption{Hyperparameters' analysis.}
	\label{fig:paras}
\end{figure}

\begin{table}[t]
\centering
\renewcommand{\arraystretch}{1.2}
\captionof{table}{The effect of MSVR on ML-1M dataset.}
\scalebox{0.95}{
\begin{tabular}{ccccc}
\toprule
\multirow{2.5}{*}{Model} & \multicolumn{4}{c}{ML-1M}       \\ \cmidrule{2-5}
& HR@5 & NDCG@5 & HR@10 & NDCG@10 \\ 
\midrule
ProtoMF        & 0.477     & 0.325       & 0.657      & 0.383        \\
ProtoMF+U-MSVR  & 0.486     & 0.330       & 0.659      & 0.388        \\
ProtoMF+I-MSVR  & 0.494     & 0.337       & 0.661      & 0.395        \\
ProtoMF+UI-MSVR & \underline{0.507}    & \underline{0.343}    & \underline{0.663}   & \underline{0.402}  \\
ProtoMF+PEO  & 0.502     & 0.341       & 0.661      & 0.397        \\
ProtoMF+RVRec & \textbf{0.519}    & \textbf{0.353}    & \textbf{0.667}   & \textbf{0.415}  \\
\bottomrule
\end{tabular}}
\label{tab:UMSVRandIMSVR}
\vspace{-0.5cm}
\end{table}

\begin{figure}[t]
    \centering
    \subfloat[\responseblack{User embeddings' visualization with TransGNN.}]{\includegraphics[width=0.243\textwidth]{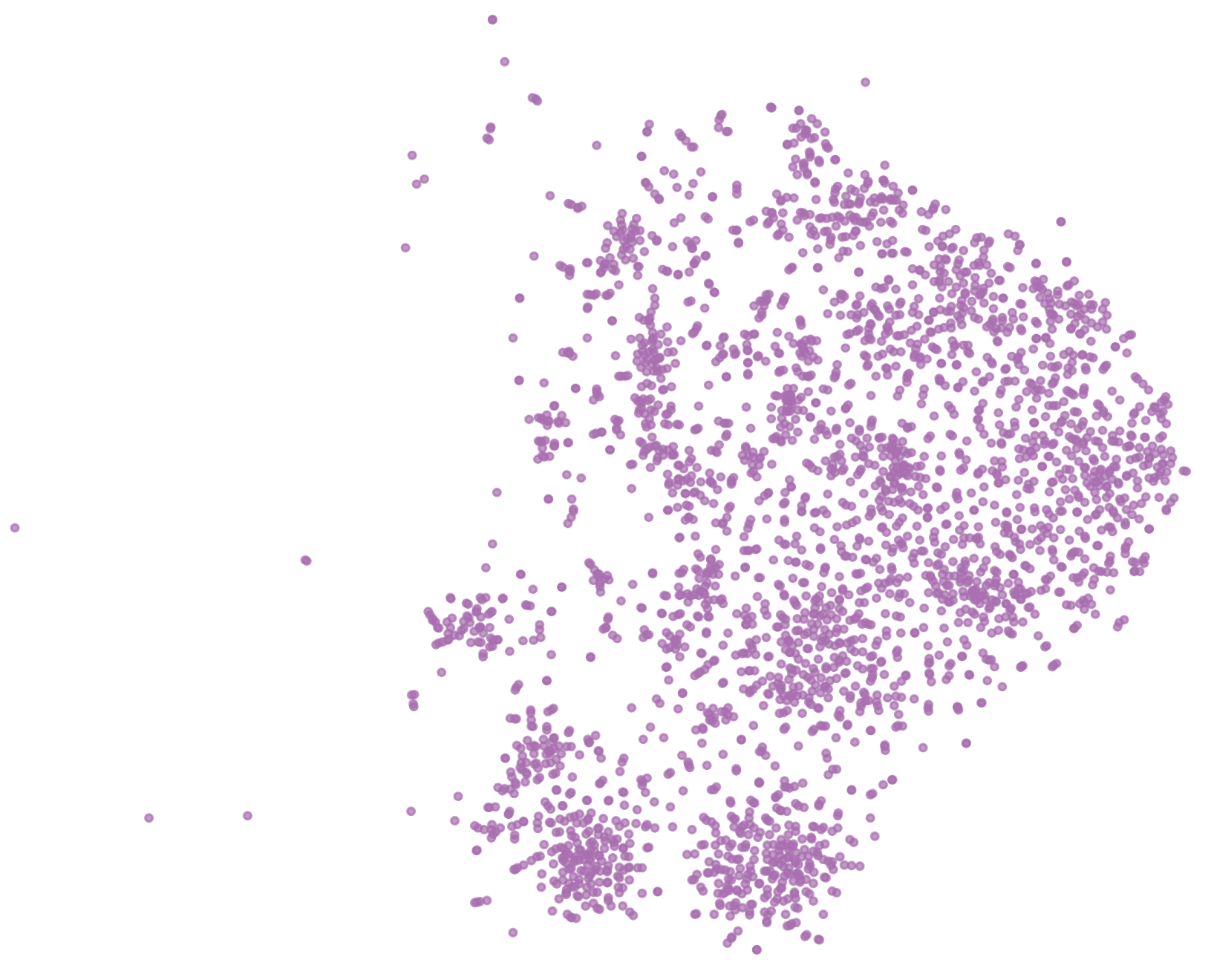}}
    \hfill 	
    \subfloat[\responseblack{Item embeddings' visualization with TransGNN.}]{\includegraphics[width=0.243\textwidth]{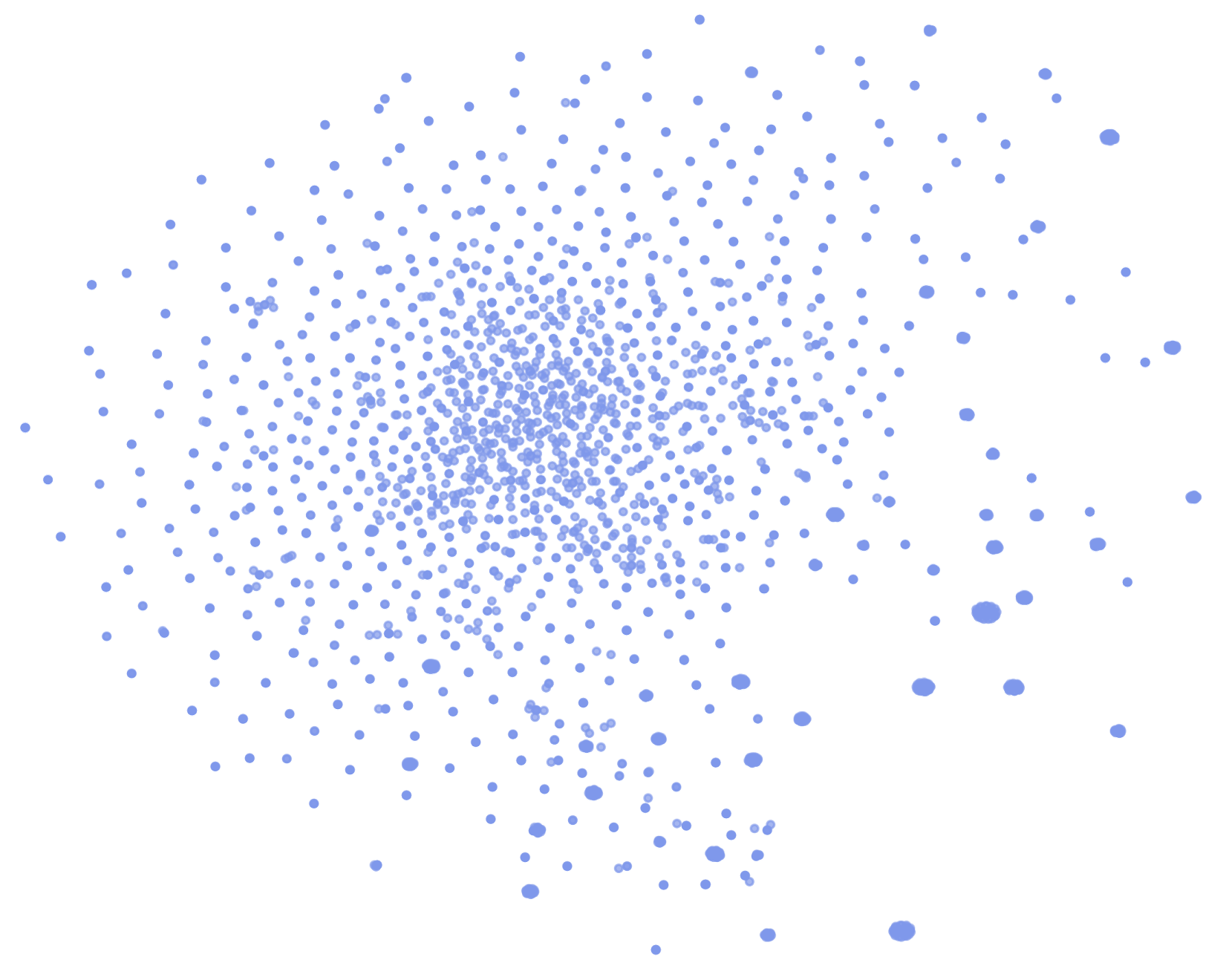}}
    \hfill
    \subfloat[\responseblack{User embeddings' visualization with TransGNN+RVRec.}]{\includegraphics[width=0.243\textwidth]{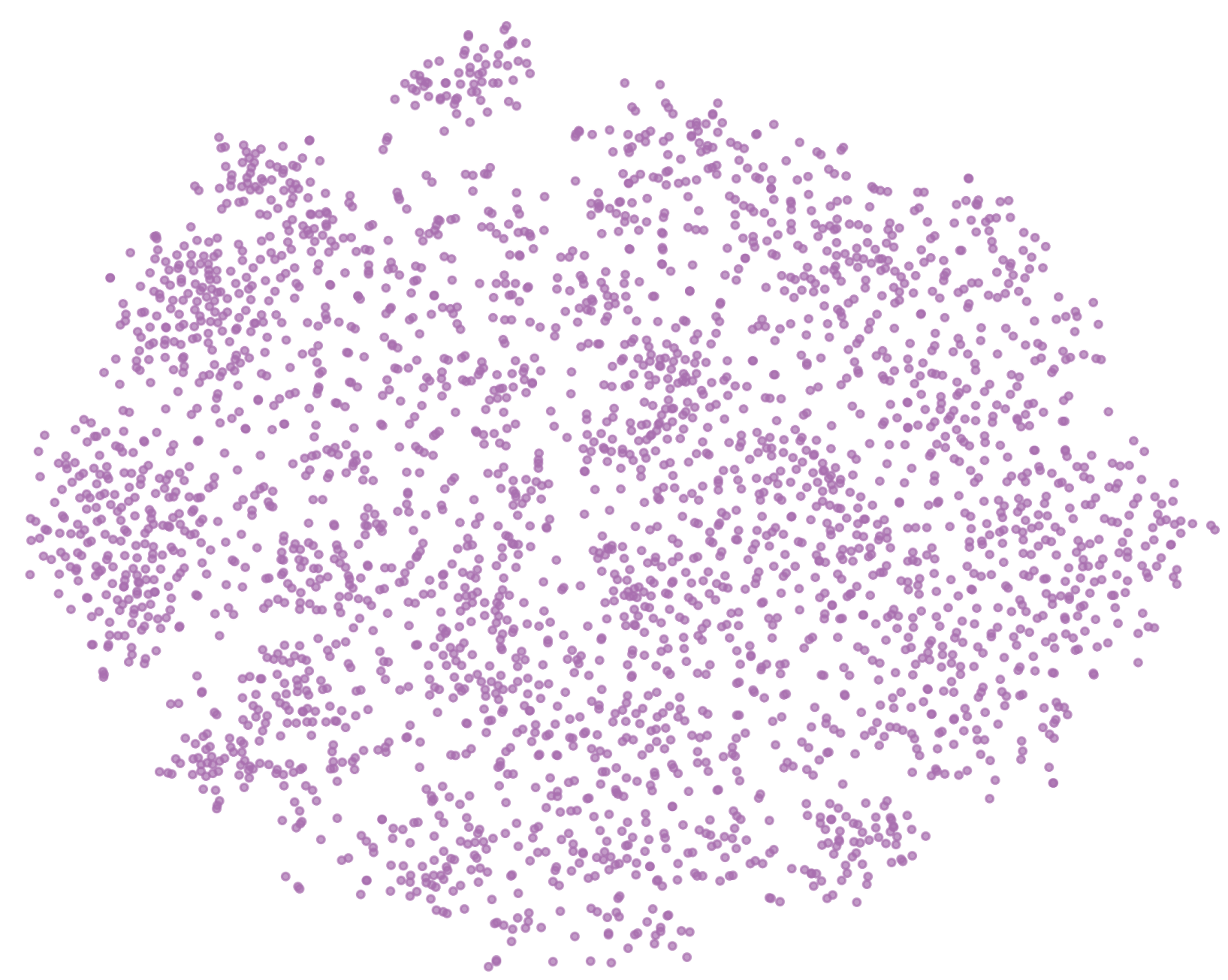}}
    \hfill
    \subfloat[\responseblack{User embeddings' visualization with TransGNN+RVRec.}]{\includegraphics[width=0.243\textwidth]{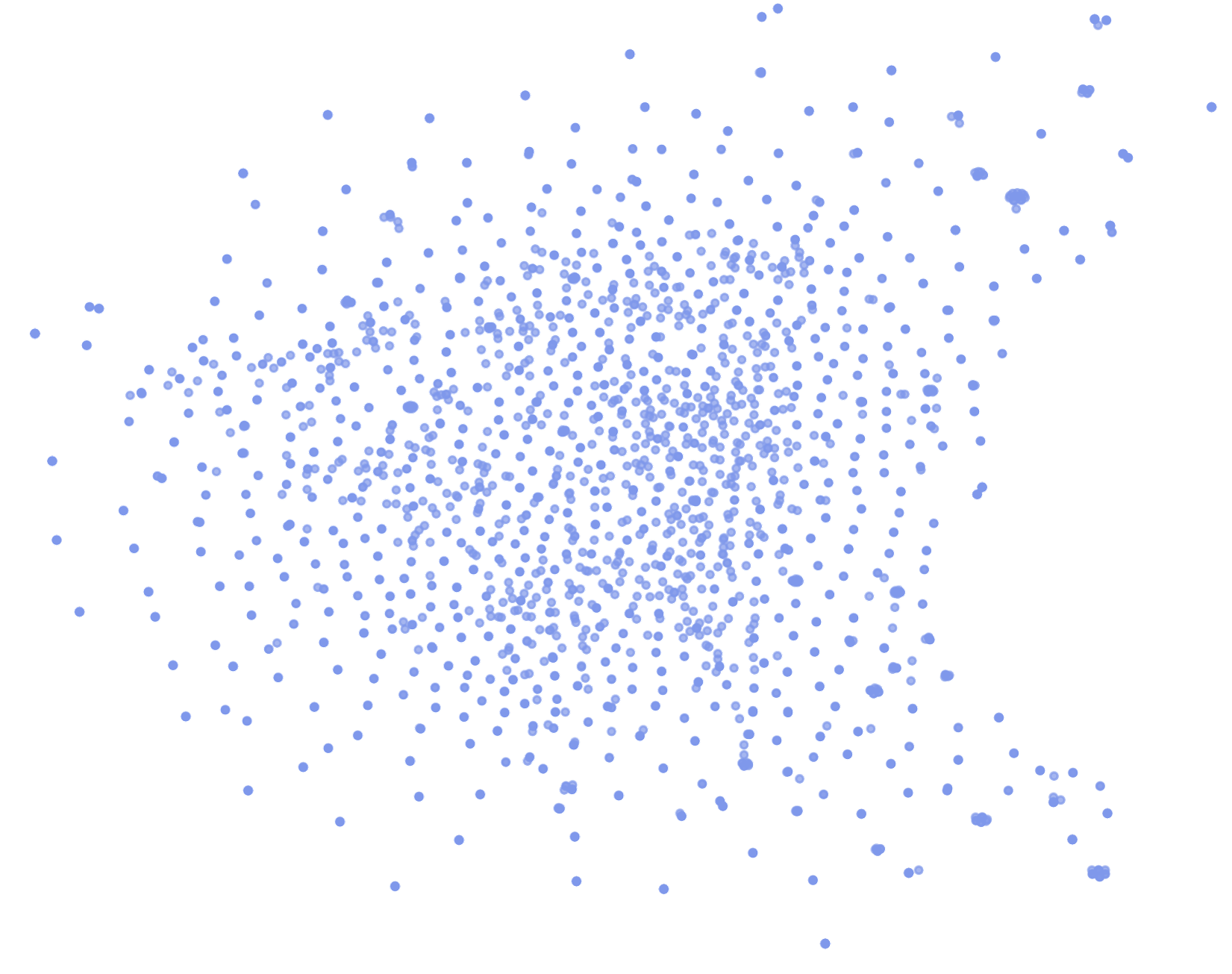}}
    \caption{\responseblack{T-SNE visualization of user and item embeddings on ML-1M dataset.}}
    \vspace{-0.5cm}
    \label{fig:tsne_visualization}
\end{figure}

\subsection{Ablation Analysis (RQ3)}
 We conduct experiments on ML-1M dataset to illustrate the impact of each module. We use MF and ProtoMF as the backbone recommender, respectively.

\subsubsection{\textbf{Effect of PEO}} In this section, \responseblack{we conduct two parts of experiments to validate the effect of PEO. Firstly, we analyze the effect of the inductive bias $\mathbf{u_i}$ in Equation (\ref{eqn:2}). We use "w/o bias" to represent PEO without $\mathbf{u_i}$, \textit{i.e.},~$\mathbf{\mu _{u_i}}= \mathbf{W_{\theta_2}}\mathbf{u_i^\ast}$. We select MF and ProtoMF as the backbones and ML-1M as the dataset. As shown in Table \ref{tab:PEObias}, incorporating the inductive bias $\mathbf{u_i}$ improves the performance across all metrics for both MF+RVRec and ProtoMF+RVRec, which demonstrate the effectiveness of the inductive bias $\mathbf{u_i}$ in Equation (\ref{eqn:2}).}

\responseblack{Then,} we compare the performance of PEO module in regular and cold-start scenarios, respectively. Following DDN, we set the cold-start scenario as an extremely sparse setting, where each user is only associated with one item for training, one for validation and one for testing. The results are shown in Figure \ref{fig:PEO}, where backbone+RVRec (w/o MSVR) represents the use of only the PEO module. Due to the capability of modelling probability distributions, we use DDN as a comparison method for further evaluation. Compared to DDN, either backbone+RVRec or backbone+RVRec (w/o MSVR) can improve the recommended performance. It suggests that the proposed PEO module can effectively capture uncertainty and improve recommendation accuracy. We observe that in the cold-start scenario, the performance of individually adding PEO for MF differs little from that of DDN. This is understandable, as DDN models embeddings as distrubutions, which perform better when there is less interaction. Meanwhile, PEO aims to obtain the optimal embeddings by modelling and optimally representing the distributions. Considering that PEO can be directly integrated into existing recommender systems, it still shows the superiority of the proposed module.

\subsubsection{\textbf{Effect of MSVR}}
In Table \ref{tab:UMSVRandIMSVR}, we compare the experimental metrics of ProtoMF, the method with only multivariate cooperative game interaction on User (ProtoMF+U-MSVR), the method with only multivariate cooperative game interaction on Item (ProtoMF+I-MSVR), the method with multivariate cooperative game interaction on both User and Item (ProtoMF+UI-MSVR) and the method only use PEO (ProtoMF+PEO) on ML-1M dataset. Firstly, we observe that ProtoMF+U-MSVR, ProtoMF+I-MSVR, and ProtoMF+UI-MSVR outperform ProtoMF indicating the effectiveness of multivariate cooperative game interactions on both User and Item in enhancing recommendation performance. Secondly, by comparing ProtoMF+UI-MSVR with ProtoMF+U-MSVR and ProtoMF+I-MSVR, we find that ProtoMF+UI-MSVR excels in all metrics, proving the effectiveness of simultaneous multivariate cooperative game interactions on both User and Item. Furthermore, ProtoMF+UI-MSVR performs better than ProtoMF+PEO, further confirming the better effectiveness of conducting multivariate cooperative game interactions. 

\subsection{\responseblack{Parameters' Analysis (RQ4)}}

In this section, we analyze two hyperparameters within the loss function to explore their impact on the model's performance. These parameters are employed to adjust the loss weights of the PEO and MSVR modules. The results are presented in Figure \ref{fig:paras}.
We systematically vary the values of $\lambda_1$ and $\lambda_2$ to investigate their influence on the model's effectiveness. Concretely, when $\lambda_1=0.001$ and $\lambda_2=1$, the model's performance resembles that of utilizing only the MSVR module. As we elevate $\lambda_1$ to 0.01 and 0.1, the model's performance gradually improves, affirming the effectiveness of the PEO module's loss. Similarly, with $\lambda_1=1$ and a gradual increase in $\lambda_2$, the model's performance consistently enhancements across all evaluation metrics, validating the effectiveness of it.

\subsection{\responseblack{Discussion on Representation Learning (RQ5)}}

\responseblack{
To provide an intuitive understanding of the learned embeddings, we randomly selected 3000 users and 3000 items from the ML-1M dataset for vector visualization. The t-SNE method was employed to reduce the high-dimensional embedding vectors into a 2-dimensional space for visualization. We select TransGNN~\cite{transgnn} as the backbone, since this method achieves best recommendation perfomance. The results are shown in Figure~\ref{fig:tsne_visualization}, where the embeddings of users and items are plotted separately for both TransGNN and TransGNN+RVRec. The embeddings learned by TransGNN+RVRec demonstrate a more balanced and comprehensive representation for both users and items, especially for users. As shown in the comparison between Figure~\ref{fig:tsne_visualization}(a) and Figure~\ref{fig:tsne_visualization}(c), the representations in Figure~\ref{fig:tsne_visualization}(a) are noticeably skewed towards the right side, while in Figure~\ref{fig:tsne_visualization}(c), the representations are distributed much more uniformly across the embedding space. A similar pattern can also be observed in the comparison between Figure~\ref{fig:tsne_visualization}(b) and Figure~\ref{fig:tsne_visualization}(d), where the latter exhibits a more balanced distribution. The results demonstrate the effectiveness of RVRec to enhance the representation learning.
}

\begin{figure}[H]
	\centering
\subfloat[\responseblack{Results on \textit{ERS-HE}.}]{\includegraphics[width=0.24\textwidth]{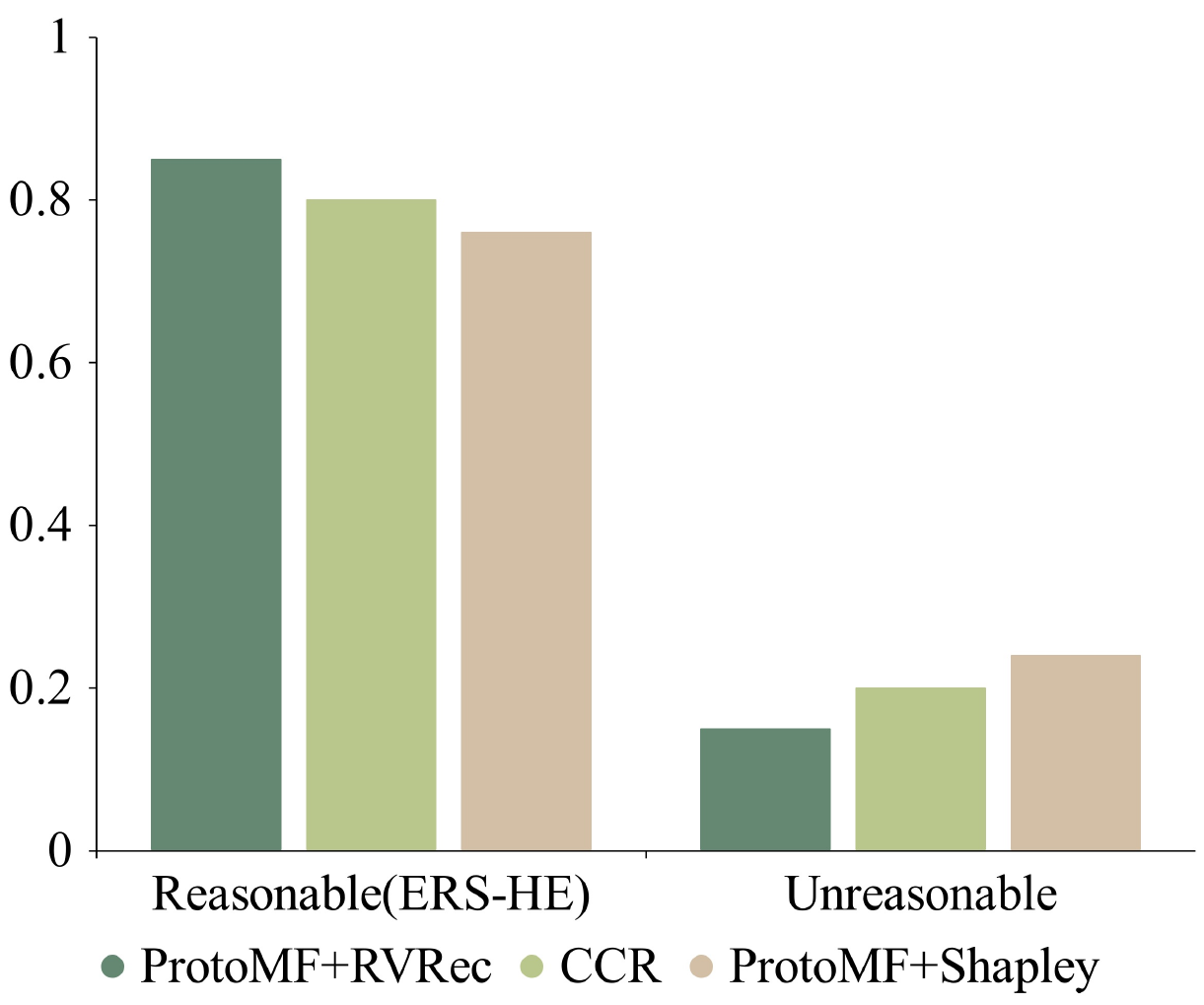}}
 \hfill 	
  \subfloat[\responseblack{Results on \textit{EQS-LE}.}]{\includegraphics[width=0.24\textwidth]{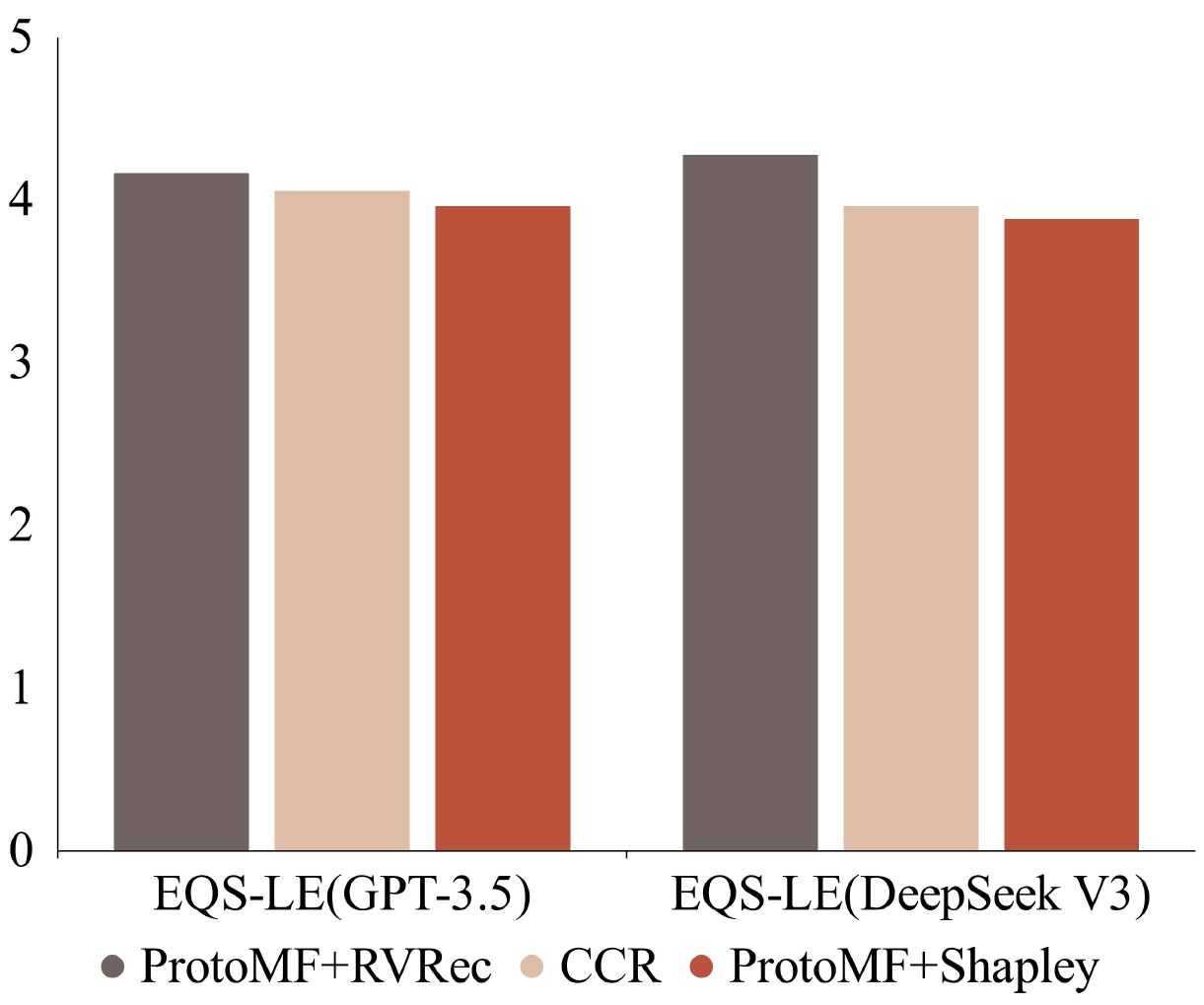}}
	\caption{\responseblack{The collected results of the explanations' evaluation by \textit{Explanations' Reasonability Score by Human Evaluator (ERS-HE)} and \textit{Explanations' Quality Score by LLM Evaluator (EQS-LE)}.}}
	\label{fig:zhongbao}
\end{figure}

\subsection{\responseblack{Explanations' Evaluation by Subjective Human Evaluator and Objective LLMs Evaluator (RQ6)}}

\responseblack{
To evaluate both the practical utility and explainability of our proposed RVRec from the perspective of user perception, inspired by DUPLE~\cite{response3}'s subjective psycho-visual test, we conduct the experiment about explanations' evaluation by subjective human evaluator and objective LLMs evaluator to judge the practical utility of the proposed RVRec effectively. For RVRec, We will use the coalition with highest multivariate Shapley value as the explanations. Specifically, we conduct two parts of explanations' evaluations by two new metrics: \textit{Explanations' Reasonability Score by Human Evaluator (ERS-HE)} and \textit{Explanations' Quality Score by LLM Evaluator (EQS-LE)}. The details for each part of explanations' evaluations and the new metrics are as follows:
}

\subsubsection{\responseblack{Explanations' evaluation with subjective human evaluator by \textit{Explanations' Reasonability Score by Human Evaluator (ERS-HE)}}}

\responseblack{
The \textit{Explanations' Reasonability Score by Human Evaluator (ERS-HE)} is designed to evaluate the subjective reasonability of explanations generated by recommender systems from a human perspective. Human evaluators judge whether the explanations align with their understanding of user preferences based on historical interactions. The metric is computed as the average proportion of explanations judged as "Reasonable" by human evaluators.
}

\responseblack{
For this evaluation, we selected 20 users from two datasets in total, with 10 users chosen from the ML-1M dataset and 10 users from the AMAZON dataset (specifically, the Video Games category). For each user, we extracted their historical interactions, averaging 28 interactions per user, along with the corresponding recommended items and explanations generated by the evaluated recommender systems: ProtoMF+RVRec, CCR, and ProtoMF+Shapley.
}

\responseblack{
We recruited 100 human volunteers (50 men and 50 women, with an average age of 30) as evaluators. Each volunteer was first presented with the selected users' historical interactions to understand their preferences. Then, they were asked to judge whether the recommended items and their corresponding explanations were "Reasonable" or "Unreasonable." The \textit{ERS-HE} score for each recommender system was calculated as the proportion of explanations judged "Reasonable" across all selected users.
}

\responseblack{
The results, shown in Figure~\ref{fig:zhongbao}(a), indicate that \textbf{85\%} of the explanations generated by ProtoMF+RVRec were judged as reasonable, outperforming CCR (\textbf{80\%}) and ProtoMF+Shapley (\textbf{76\%}). These results highlight the effectiveness of ProtoMF+RVRec in providing user-friendly and explainable recommendations. However, \textbf{15\%} of the evaluators expressed uncertainty about certain explanations. This may stem from implicit relationships between items that are not intuitive. For instance, when historical items include iPhones, AirPods, and other electronic devices, our methodology assumes that the removal of a single item (\textit{e.g.}, iPhones) has a more significant impact than removing multiple items together. While this assumption often holds, it may fail for users who prefer iPhones but not AirPods, leading to less convincing explanations for certain individuals. These findings highlight the strengths of ProtoMF+RVRec in generating reasonable explanations for the majority of users, while also revealing areas for improvement in addressing less intuitive item relationships.
}

\subsubsection{\responseblack{Explanations' evaluation with objective LLMs evaluator by \textit{Explanations' Quality Score by LLM Evaluator (EQS-LE)}}}

\responseblack{
The \textit{Explanations' Quality Score by LLM Evaluator (EQS-LE)} is designed to evaluate the quality of explanations generated by recommender systems from an objective perspective using large language models (LLMs). The quality of each explanation is assessed based on three criteria: logical consistency, linguistic clarity, and relevance to user preferences. LLMs assign a score ranging from \textbf{1} (lowest) to \textbf{5} (highest) for each explanation, and the final \textit{EQS-LE} score is computed as the average score across all explanations.
}

\responseblack{
To conduct this evaluation, we selected 100 users from two datasets in total, with 50 users from the ML-1M dataset and 50 users from the AMAZON dataset (specifically, the Video Games category). For each user, we extracted their historical interactions, averaging 36 interactions per user, along with the corresponding recommended items and explanations generated by the evaluated recommender systems: ProtoMF+RVRec, CCR, and ProtoMF+Shapley.
}

\responseblack{
We utilized two state-of-the-art LLMs, GPT-3.5 and DeepSeek V3, as evaluators. The LLMs were provided with the selected users' historical interactions, the corresponding recommended items and the explanations generated by ProtoMF+RVRec, CCR, and ProtoMF+Shapley. Each explanation was evaluated and assigned a score based on the aforementioned criteria. The final \textit{EQS-LE} score for method was calculated as the average score across all explanations.
}

\responseblack{
The results, shown in Figure~\ref{fig:zhongbao}(b), indicate that ProtoMF+RVRec achieved the highest average scores, with \textbf{4.16} from GPT-3.5 and \textbf{4.27} from DeepSeek, outperforming CCR (\textbf{4.05} by GPT-3.5 and \textbf{4.02} by DeepSeek) and ProtoMF+Shapley (\textbf{3.96} by GPT-3.5 and \textbf{3.88} by DeepSeek). These results demonstrate that ProtoMF+RVRec generates explanations of superior quality, particularly in terms of logical consistency and linguistic clarity, compared to the baseline methods.
}

\section{Conclusion}

In this work, we propose RVRec, a plug-and-play model-agnostic embedding enhancement approach, which can pipe into existing recommender systems to improve both explainability and personality. We focus on two problems of the prototype in recommender systems, \textbf{suboptimal representation} and \textbf{low-value}. By introducing a probability-based embedding optimization method that uses a contrastive loss based on negative 2-Wasserstein distance, RVRec is aimed to enhance the representativeness of the embeddings. What's more, we introduce a reweighing method based on game theory and multivariate Shapley values strategy proposed by us, aiming to evaluate and explore the value of interactions and embeddings. Extensive experiments show that RVRec performs competitively against state-of-the-art baselines, with significant improvement both in personalization and explanation.

\vfill

\end{document}